\documentclass[aps,pra,superscriptaddress,twocolumn,nofootonbib,floatfix,longbibliography]{revtex4-2} 

\usepackage[final]{graphicx}    
\usepackage{soul}
\usepackage[dvipsnames]{xcolor}
\usepackage[utf8]{inputenc}
\usepackage[T1]{fontenc}
\usepackage{ulem}
\usepackage{natbib}
\usepackage{framed}
\usepackage{amsfonts}
\usepackage{amsmath}
\usepackage{mathtools}
\usepackage{amssymb}
\usepackage{hyperref}
\hypersetup{colorlinks=true}

\newcommand{\newchange}[1]{#1}

\newcommand{\change}[1]{#1}

\newcommand{\Ket}[1]{\left|#1\right>}
\newcommand{\Bra}[1]{\left<#1\right|}

\newcommand{\trace}[1]{\mathrm{Tr}\left(#1\right)}

\DeclarePairedDelimiter\floor{\lfloor}{\rfloor}

\begin{document}

\title{Quantum ergodicity and scrambling in quantum annealers}


\author{Manuel H. Muñoz-Arias}
\email{munm2002@usherbrooke.ca}
\affiliation{Institut Quantique and Département de Physique, Université de Sherbrooke, Sherbrooke, Quebec, J1K 2R1, Canada}
\altaffiliation{Peresent address: Quantum Algorithms and Applications Collaboratory, Sandia National Laboratories, Livermore, CA 94550, USA}
\author{Pablo M. Poggi}
\email{pablo.poggi@strath.ac.uk}
\altaffiliation{Corresponding author.}
\affiliation{Department of Physics, SUPA and University of Strathclyde, Glasgow G4 0NG, United Kingdom}

\date{\today}
\begin{abstract}
    Quantum annealers play a major role in the ongoing development of quantum information processing and in the advent of quantum technologies. Their functioning is underpinned by the many-body adiabatic evolution connecting the ground state of a simple system to that of an interacting classical Hamiltonian which encodes the solution to an optimization problem. Here we explore more general properties of the dynamics of quantum annealers, going beyond the low-energy regime. We show that the unitary evolution operator describing the complete dynamics is typically highly quantum chaotic. As a result, the annealing dynamics naturally leads to volume-law entangled random-like states when the initial configuration is rotated away from the low-energy subspace. Furthermore, we observe that the Heisenberg dynamics of a quantum annealer leads to extensive operator spreading, a hallmark of quantum information scrambling. \newchange{In contrast, we find that when the annealing schedule is returned to the initial configuration (i.e. via a cyclic ramp), a subtle interplay between chaos and adiabaticity emerges, and the dynamics shows strong deviations from full ergodicity}.

\end{abstract}

\maketitle

\section{Introduction}
\label{sec:introduction}
The goal of exploiting quantum effects to solve computational problems more efficiently than classical devices has led to the fast-paced development of quantum technologies. A remarkably simple, yet potentially powerful, approach is quantum annealing~\cite{Rajak2023}, which uses adiabatic quantum evolution to tackle problems in combinatorial optimization. More broadly, adiabatic dynamics can be leveraged to perform universal quantum computation~\cite{Albash2018}, in quantum control protocols for entangled-state preparation~\cite{vitanov2017}, and is closely related to variational quantum algorithms like the quantum approximate optimization algorithm (QAOA)~\cite{Farhi2014,Blekos2024}. 

\begin{figure}[t!]
\centering
\includegraphics[width=0.95\columnwidth]{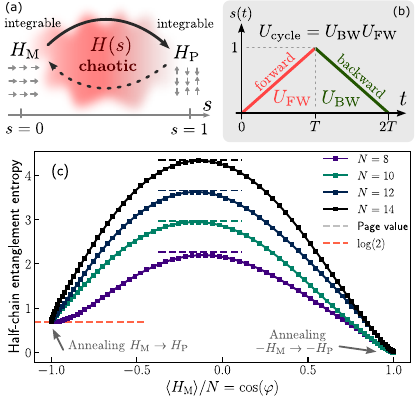}
\caption{(a) Schematic of a quantum annealing protocol. Two integrable Hamiltonians $H_{\rm M}$ and $H_{\rm P}$ are interpolated by chaotic Hamiltonian $H(s)$. Slowly varying the parameter $s$ from $0$ to $1$ will connect the ground state of $H_{\rm M}$ to that the ground state of $H_{\rm P}$. (b) The forward ramp from $s(t=0)=0$ to $s(t=T)=1$ leads to the unitary map $U(T)\equiv U_{\rm FW}$. Ramping back from $s=1$ to $s=0$ generates the backwards unitary $U_{\rm BW}$ such that the cyclic process is given by $U_{\rm cycle}=U_{\rm BW}U_{\rm FW}$. (c) Half-chain entanglement entropy of the state $\Ket{\psi(T)}$ at the end of a forward ramp under annealing dynamics, for the Ising model described in Sec.~\ref{sec:model}. The initial state is the spin coherent state $\Ket{\psi(0)} =  \Ket{\theta=0,\varphi}$, where all qubits are pointing along the direction $(\cos(\varphi),\sin(\varphi),0)$ on the Bloch sphere. For $\varphi\rightarrow 0,\:\pi$, the dynamics connects the ground state of $\pm H_{\rm M}$ to that of $\pm H_{\rm P}$ and generate little entanglement. For intermediate values of $\varphi$, the final states acquire volume-law entanglement entropy, similar to that of random states. }
\label{fig:fig1}
\end{figure}

From a many-body physics perspective, the richness of adiabatic dynamics arises from the interplay between interactions and time-dependent driving. Interestingly, generic quantum many-body systems, in absence of extensive symmetries, are typically quantum chaotic~\cite{Rigol2008,Polkovnikov2011,Santos2010,Dalessio2016,Munoz2022}, displaying features such as random-matrix-like spectral statistics~\cite{Dalessio2016}, pseudorandom eigenstates~\cite{Gubin2012,Lydzba2021,Kliczkowski2023}, fast dynamical generation of entanglement~\cite{wang2004}, and three-stage dynamics in the entanglement spectrum~\cite{Chang2019}. The addition of driving and the ensuing lack of energy conservation typically enhances the dynamical signatures of chaos~\cite{Dalessio2014}, and can turn integrable systems into chaotic ones~\cite{haake1987}. 

In this context, adiabatic driving can be regarded as the ``gentlest'' form of time-dependence, as the slow change in system parameters leads to a quasi-static connection between eigenstates of different Hamiltonians. In generic many-body systems, however, such connection can only realistically be achieved in the low-energy regime, where states can be protected by a sufficiently large energy gap. This suffices to guarantee the functioning of a quantum annealer, which focuses on ground states. But, the dynamical properties of these systems at other energy scales will be determined by a nontrivial interplay between driving, (non)adiabaticity, and chaos. So far, this regime has remained largely unexplored in the literature.

In this paper we explore many-body adiabatic evolution beyond the low-energy sector. We consider a system of spin-$\frac{1}{2}$ particles whose Hamiltonian is slowly interpolated between two integrable models through a chaotic path. This is the typical scenario in quantum annealing, where the system is driven from a ``mixer'' (non-interacting) Hamiltonian $H_{\rm M}$ at $t=0$ to a ``problem'' (classical) Hamiltonian $H_{\rm P}$ at $t=T$, as depicted in Fig.~\ref{fig:fig1}a and~\ref{fig:fig1}b. We first explore the dynamics of the system when the initial product state is rotated away from the ground state of $H_{\rm M}$, by studying the entanglement properties of the final state after the ``anneal'', $\Ket{\psi(T)}$. Then, we probe global properties of the adiabatic evolution by analyzing properties of the unitary operator $U(t)$ dictating the full dynamics, for which we use standard diagnostics of quantum chaos based on the spectrum of eigenphases and eigenvectors of $U(t)$. Finally, we study the dynamics of generic operators in the Heisenberg picture through the formalism of operator growth and demonstrate the onset of quantum information scrambling. 

The emergence of chaotic properties in the dynamics of a quantum annealer can be readily observed in a simple example shown in Fig. \ref{fig:fig1}c. There, we plot the half-chain entanglement entropy of the state $\Ket{\psi(T)}$ at the end of the annealing schedule, as a function of the mean energy $\langle H_{\mathrm{M}}\rangle$ of the initial product state for different system sizes $N$. For $\langle H_{\mathrm{M}}\rangle=\pm N$, the initial state is the ground state of $\pm H_{\rm M}$, i.e. the standard annealing scenario, and in consequence we obtain that the corresponding final states are only slightly entangled. However, as the initial state is rotated away from the ground state (by angle $\varphi$, as described in Sect. \ref{sec:evol_scs}) we find that the final configuration acquires an entanglement entropy which grows with system size. At the peak, the final states exhibit clear volume-law entanglement entropy, well approximated by Page's value~\cite{Page1993}. We thus observe that the annealing dynamics is generating random-like states with very high entanglement, akin to what is expected from the time evolution of a fully quantum chaotic system. 

As we show in this paper, the observed chaotic features turn out to be generic and emerge in the dynamics of most initial states. Interestingly, we find that significant deviations from \newchange{full ergodicity} are observed if one considers a cyclic annealing schedule, where the interpolation parameter $s(t)$ is driven from $s(t=0)=0$ to $s(t=T)=1$ and then back to $s(t=2T)=0$; see Fig. \ref{fig:fig1} (b). For such case, we observe the existence of eigenstates of $U(2T)$ that show clear signatures of low-energy states, in particular displaying area-law entanglement, and that the dynamics of global operators show unscrambling features \newchange{(i.e. operator growth followed by operator shrinking). Thus, in this regime we find that driving does not enhance ergodicity, but rather leads to a subtle interplay where adiabatic and chaotic features coexist.}

The rest of the manuscript is organized as follows. In Sec.~\ref{sec:model} we introduce the model we use in this paper, characterize the onset of chaos and integrability for different parameter values, and discuss the connection between adiabaticity and the spectral gaps throughout the energy spectrum. In Sec.~\ref{sec:evol_scs} we analyze the evolution of initial states that are rotated away from the ground state of $H_\mathrm{M}$, expanding on the example discussed in Fig.~\ref{fig:fig1} and considering both forward and cyclic adiabatic ramps. In Sec.~\ref{sec:chaos_in_unitary} we study properties of the many-body time-evolution operator $U(t)$ describing the dynamics. We focus on the analysis of the statistical properties of its eigenphases and entanglement structure of its eigenstates. We connect the behavior of these quantities to the emergence of chaotic signatures in the system for both forward and cyclic driving protocols. In Sec.~\ref{sec:scrambling} we study the adiabatic dynamics of the system in the Heisenberg picture. For that, we consider different sets of initial operators and probe the emergence of quantum information scrambling by tracking how the mean operator size evolves in time for the two driving protocols. Finally, in Sec.~\ref{sec:discussion_outlook} we summarize our findings, discuss how our results could inform new uses and applications of quantum annealers, and propose potential avenues for future work.

\section{Physical model and driving protocol}
\label{sec:model}

We consider a system of $N$ spin-$\frac{1}{2}$ particles described by a Hamiltonian of the form
\begin{equation}
    H(s) = s H_{\rm M} + (1-s) H_{\rm P},
    \label{eq:hami_s}
\end{equation}
where 
\begin{align}
    H_{\rm M} &= \sum\limits_{i=1}^N \sigma_i^x \\
    H_{\rm P} &= \sum\limits_{i,j=1}^{N} \chi_{ij} \sigma_i^z \sigma_{j}^z + \sum\limits_{i=1}^N \lambda_i \sigma_i^z
\end{align}
and where $\{\sigma_i^\alpha\}$ with $\alpha=x,y,z$ denote the usual Pauli operators acting on site $i=1,\ldots,N$. The parameter $s\in [0,1]$ allows to interpolate between the noninteracting mixer Hamiltonian $H_{\rm M}$, and the interacting classical Hamiltonian $H_{\rm P}$ which is diagonal in the computational basis for any choice of connectivity $\{\chi_{ij}\}$ and local longitudinal fields $\{\lambda_i\}$. The problem Hamiltonian $H_{\rm P}$ can encode instances of quadratic binary optimization (QUBO) problems~\cite{Lucas2014} for instance maximum cut and maximum independent set \cite{Ebadi2022}, as well as describe physical models of relevance to the theory of spin glasses and statistical mechanics~\cite{Sherrington1975,Charbonneau2023}. In this work we focus entirely on the simplest scenario, where interactions are nearest-neighbor $\chi_{ij}=\delta_{j,i+1}$ (with open boundary conditions) and the local field is homogeneous across all sites  $\lambda_i=\lambda=1$. This choice makes $H(s)$ in Eq.~(\ref{eq:hami_s}) equivalent to the mixed-field Ising model~\cite{Karthik2007,Kim2013,Kim2015}.

\begin{figure}
\centering
\includegraphics[width=0.97\columnwidth]{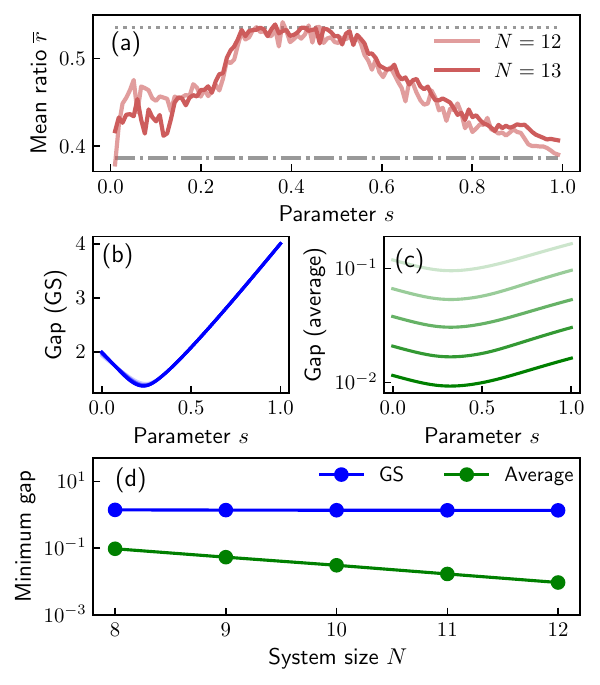}
\caption{Spectral properties of Hamiltonian $H(s)$ in Eq. (\ref{eq:hami_s}). (a) Mean level spacing ratio $\overline{r}$, defined from Eq. (\ref{eq:ratios_def}), as a function of the interpolation parameter $s$. Average is taken over the bulk of the spectrum, here defined by not considering the top 5\% lowest and highest energy levels. Dotted (dash-dotted) lines indicate the values of $\overline{r}$ for chaotic (GOE) and integrable (Poisson) models. (b) and (c) Spectral gaps as a function of $s$ for system sizes $N\in [8,12]$ (light to dark tones). Figure (b) shows the ground state gap $\Delta_0(s)$, while (c) shows the average gap $\sum_j \Delta_j/(\overline{d}-1)$ with $\overline{d}$ the relevant Hilbert space dimension. (d) Scaling of the minimum ground state gap and average gap with system size $N$; minimum is taken over $s\in[0,1]$. All calculations are performed in the positive parity sector of the Hamiltonian $H(s)$, whose dimension is $d=2^{N-1}$.}
\label{fig:figure2}
\end{figure}

The Hamiltonian in Eq.~(\ref{eq:hami_s}) is trivially integrable for $s=0$ and $s=1$, while for $s\in (0,1)$ integrability is broken and the model becomes quantum chaotic. As discussed in Ref.~\cite{grabarits2024}, a similar situation is found for most choices of connectivity graphs $\chi_{ij}$ (at least, those far from the disconnected and completely connected regimes). This motivates our choice of the nearest-neighbor graph as a simple representative of the typical case leading to a chaotic path interpolating between two integrable limits, as depicted in Fig.~\ref{fig:fig1}a. The onset of chaos can be diagnosed by the emergence of level repulsion in the bulk of the energy spectrum of the Hamiltonian, which in turn can be quantified by the mean level spacing ratio (MLSR) $\overline{r}$~\cite{Atas2013}. Given a set of eigenvalues $\{e_i\}$, the associated level spacings (gaps) are
\begin{equation}
    \Delta_j= e_{j+1}-e_j
    \label{eq:gaps_def}
\end{equation}
and we define the spacing ratios as
\begin{equation}
    r_j = \frac{\max (\Delta_j,\Delta_{j+1})}{\min (\Delta_j,\Delta_{j+1})}.
    \label{eq:ratios_def}
\end{equation}
The MLSR is computed as the mean of the set $\{r_j\}$ across the bulk of the spectrum. For quantum chaotic Hamiltonians, the MLSR matches random matrix theory predictions; for the case of the model considered here, the relevant matrix ensemble is the Gaussian Orthogonal Ensemble for which $\overline{r}_{\rm GOE}\simeq  0.535$~\cite{Atas2013}. Integrable systems show uncorrelated level statistics yielding $\overline{r}_{\rm INT} \simeq 0.386$~\cite{Atas2013}. Importantly, for the MLSR (and any other spectral property) to be meaningful one must consider a given symmetry sector of the Hamiltonian. For the model in Eq.~(\ref{eq:hami_s}), the only discrete symmetry present is related to the invariance of the system when performing a reflection with respect to the middle of the chain. We choose to work in a positive parity sector since this is where the ground state of $H_{\rm M}$ resides.

In Fig.~\ref{fig:figure2} (a) we plot the MLSR $\overline{r}$ for the Hamiltonian in Eq.~(\ref{eq:hami_s}) as a function of the interpolation parameter $s$ for the chosen geometry. Dotted (dash-doted) lines indicate expected limiting values for chaotic (integrable) models. We observe that the model is maximally chaotic for $s\simeq 0.3-0.6$, with clear deviations from the GOE predictions as $s=0$ and $s=1$ are approached.  We emphasize that this characterization of quantum chaos corresponds to the static Hamiltonian $H(s)$ where $s$ is fixed. Nonetheless, these results will inform our analysis of the properties of the adiabatically-driven system in the next sections.

\subsection{Gap characterization and adiabaticity}

We are interested in analyzing the dynamics of the system when the parameter $s=s(t)$ is slowly changed in time. We consider two protocols: a standard \textit{forward} ramp, and a \textit{cyclic} ramp. In the forward ramp, $s(t)=t/T$ and the Hamiltonian is interpolated between $H_{\rm M}$ at $t=0$ to $H_{\rm P}$ at $t=T$. This is the usual scenario in quantum annealing where one expects to connect the ground state of $H_{\rm M}$ to that of $H_{\rm P}$. In the cyclic ramp, on the other hand, $s(t)=t/T$ for $t\in[0,T]$ and $s(t)=(2T-t)/T$ for $t\in (T,2T]$, such that the Hamiltonian returns to $H_M$ at the final time $t=2T$. In this case, the system initially prepared in one of the eigenstates of $H_{\rm M}$ will return to the same eigenstate at $t=2T$ if the evolution is truly adiabatic, having acquired an overall geometric phase in the process \cite{Berry1984}. 

A key aspect of designing the adiabatic driving is to make sure that the conditions of the Adiabatic Theorem are met throughout the evolution. For $H(s)=H(s(t))$ and $s\sim t/T$, these can be roughly translated into the requirement~\cite{Born1928,Kato1950}.
\begin{equation}
    T \gg \max_{s\in [0,1]}  \frac{\lvert\Bra{k} \frac{\partial H}{\partial s}\Ket{n}\rvert}{\Delta_n(s)^2}
    \label{eq:adiabatic_condition}
\end{equation}
where $\Ket{n}$ and $\Ket{k}$ denote instantaneous eigenstates of $H(s)$ and $\Delta_n(s)$ is the spectral gap of Eq.~(\ref{eq:gaps_def}). In a generic many-body system, however, the exponential amount of energy levels in a given fixed energy window (in the bulk of the spectrum) precludes the validity of Eq.~(\ref{eq:adiabatic_condition}) for the entirety of the spectrum even for moderate system sizes. For the model we consider this can be seen by analyzing the scaling of the spectral gaps $\Delta_j(s)$ defined in Eq.~(\ref{eq:gaps_def}). In Fig.~\ref{fig:figure2}b and \ref{fig:figure2}c we plot the ground state gap $\Delta_0(s)$ and the average gap 
\begin{equation}
    \overline{\Delta}(s) = \frac{1}{\overline{d}-1} \sum\limits_{j=1}^{\overline{d}} \Delta_j(s),
\end{equation}
where $\overline{d}$ indicates the dimension of the positive parity subspace, $\overline{d}\simeq 2^{N-1}$. Curves are shown for various system sizes $N\in [8,12]$, revealing that the ground state gap remains finite as the system is scaled up, while the average gap shrinks exponentially. This is confirmed by the data in Fig.~\ref{fig:figure2} (d) which shows the scaling of the minimum values of $\Delta_0(s)$ and $\overline{\Delta}(s)$ with system size. These results show that in order to strictly meet the adiabaticity condition Eq.~(\ref{eq:adiabatic_condition}) for every energy level, the evolution time would have to grow exponentially with system size, making it practically unfeasible. 

This effective lack of adiabaticity in a many-body system is not surprising, and it can be seen as a consequence of the exponential growth of Hilbert space dimension (nonetheless, see Ref.~\cite{Yarloo2024,Puebla2024} for important exceptions). Here we take the approach of considering evolution times $T$ that are long enough to meet condition (\ref{eq:adiabatic_condition}) for the low (and high) energy portions of the spectrum - in particular for the ground state of the system. These are the conditions in which quantum annealers are ideally designed to work in. Our focus will be in analyzing the dynamical process \textit{as a whole}, without restricting to analyzing only the evolution of the ground state. 

\section{Evolution of spin coherent states}
\label{sec:evol_scs}

In this section we explore the dynamics of different states under the slow adiabatic driving. Formally, the annealing dynamics is described by the set of unitaries
\begin{equation}
    U(T_{\rm tot}) \coloneq \mathcal{T}\left[ e^{-i\int_{0}^{T_{\rm tot}} dt H(s(t))} \right]
\end{equation}
with $\mathcal{T}[.]$ the time ordering operator, $T_{\rm tot} = T,\: 2T$ for the forward and cyclic ramps, respectively, and $s(t)$ chosen accordingly. In our numerical simulations we approximate the ideal protocol with a discretized version of the form 
\begin{equation}
\label{eq:approx_adiabatic_unitary}
U(T_{\rm tot}) \approx \prod_{m=0}^{q-1} e^{-i m\Delta t H(s(m\Delta t))},
\end{equation}
with $\Delta t$ the time step and $q = T_{\rm tot}/\Delta t$ the total number of steps. Hence, the protocol is controlled by the two independent parameters $(\Delta t, T_{\rm tot})$, which can always be chosen to guarantee the fulfillment of the adiabatic condition (for the low energy sector). Here, we take $T_{\rm tot} \propto N^2$~\cite{Jansen2007,Schiffer2024} and $\Delta t < 1$~\cite{Yi2021,Kovalsky2023}, which satisfy the above requirements.  

We begin our exploration of the quantum ergodic properties in the annealing process by considering the evolution of product states where initially all spins are polarized along the same direction $\vec{n}$, such that
\begin{equation}
  |\psi(0)\rangle = \lvert\uparrow_{\vec{n}}\rangle^{\otimes N} = |\theta, \varphi\rangle 
\end{equation}
are spin coherent states (SCS)~\cite{Radcliffe1971}, where $\vec{n}$ is defined by the polar and azimuthal angles $(\theta, \varphi)$. Note that $|\pi/2, \pi\rangle = |\psi(0)\rangle = |-\rangle^{\otimes N}$ and $|\pi/2, 0\rangle = |\psi(0)\rangle = |+\rangle^{\otimes N}$ are the lowest and highest energy eigenstates of $H_{\rm M}$, respectively. We define a continuous family of SCSs connecting these two extremal states as
\begin{equation}
    \mathcal{S} = \{|\pi/2, \varphi\rangle\:  |\:   \varphi\in[0, \pi]\},
\end{equation}
which cover the full range of energies of $H_{\rm M}$ since $\langle H_{\rm M} \rangle = \langle \pi/2,\varphi| H_{\rm M} | \pi/2, \varphi\rangle = N\cos(\varphi)$.  For the results discussed in this section we always take $|\psi(0)\rangle \in \mathcal{S}$.

For both the forward and cyclic protocols we investigate the ergodicity of the evolution by computing the half-partition entanglement entropy of the final state. This quantity measures the degree of entanglement between two subsets $A$ and $B$ in a collection of $N$ spins, with $|A\cup B| = N$. Given a pure state $|\psi\rangle$ of the $N$ spins, the reduced density matrix of the subsystem $A$ (respectively, $B$) $\rho_{A} = {\rm tr}_{B}(|\psi\rangle \langle \psi|)$ associated with the considered bipartition gives access to the half-partition entanglement entropy as the von Neumann entropy of the reduced state
\begin{equation}
\label{eq:half_chain_ee}
S_{A} = -{\rm tr}\left(\rho_{A} \log(\rho_A)\right) = -\sum_{l}\lambda_l \log(\lambda_l),
\end{equation}
where $\{\lambda_l\}_{l=1,..,2^{|A|}}$ are the eigenvalues of $\rho_{A}$ satisfying $\sum_{l}\lambda_l = 1$. For the rest of the manuscript we fix $A$ as the block of the $N/2$ left-most spins in the chain and $B$ the complement, and thus we will refer to the quantity in Eq.~(\ref{eq:half_chain_ee}) as half-chain entanglement entropy. Typically, time evolution under a time-independent chaotic Hamiltonian or chaotic unitary transforms initial product states, with $S_{A}=0$, into highly entangled states which are indistinguishable from random pure states with $S_{A}$ saturating the Page value $S_A \sim \frac{\log(2)}{2}N$~\cite{Page1993}.

\begin{figure}
\includegraphics[width=0.98\linewidth]{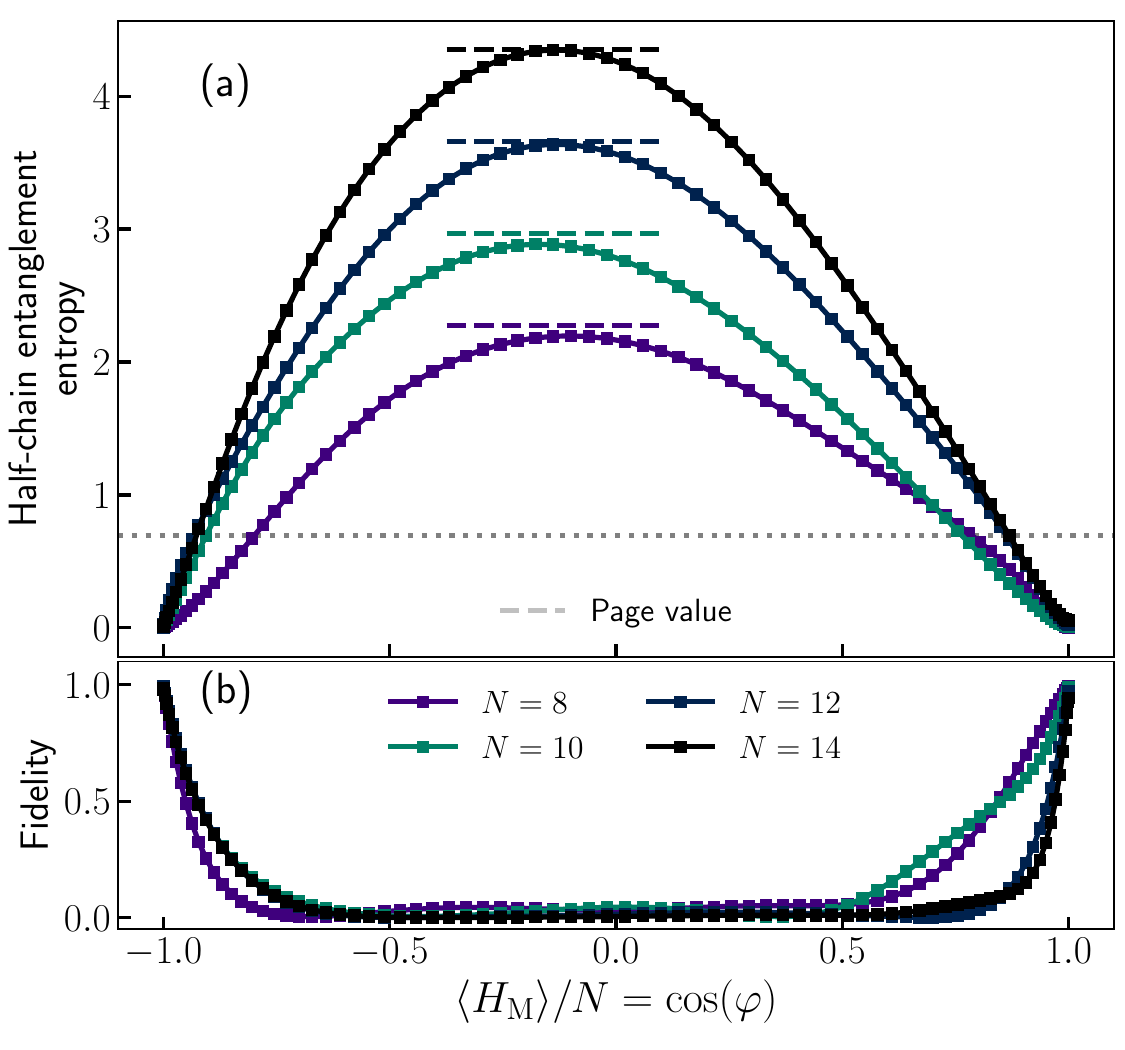}
\caption{Properties of the final state, $|\psi(2T)\rangle$, for a cyclic ramp initialized with a spin coherent state pointing along the direction given by angles $(\theta, \varphi) = (\pi/2, \varphi)$ (see main text for details), for systems $N=8,10,12,14$. (a) Half-chain entanglement entropy $S_{A}$ of $|\psi(2T)\rangle$ as a function of the mean-energy density of the initial state which is determined by its direction. The dashed lines indicate the Page value for the respective system size. The dotted line indicates $S_{A} = \log(2)$. (b) Fidelity between initial and final states of the cyclic ramp as function of the initial state mean-energy density. A perfect cyclic evolution is characterized by a unit fidelity. Only states with a small mean-energy density neighborhood of either $|+\rangle^{\otimes N}$ and $|-\rangle^{\otimes N}$ can be successfully returned to their initial configuration. Parameters are: $T=600$, $\Delta t = 0.05$.}
\label{fig:figure_scs}
\end{figure}

\subsection{Forward ramps}
\label{subsec:scs_forward}
First, we consider the evolution of every state $|\psi(0)\rangle \in \mathcal{S}$ with a forward ramp and compute $S_{A}$ of the final state $|\psi(T)\rangle$. When $|\psi(0)\rangle = |\pi/2, \pi\rangle$ with $\langle H_{\rm M}\rangle/N = 1$, this is the standard quantum annealing scenario. Naturally, the ground state of $H_{\rm P}$, \change{given by (for $N$ even and open boundary conditions) $|{\rm GS}\rangle = \frac{1}{\sqrt{2}}( \lvert\uparrow\downarrow\rangle^{\otimes N/2} \pm  \lvert\downarrow\uparrow\rangle^{\otimes N/2})$ with energy $E_{\rm GS} = -(N-1)$}, is reached at the end of the ramp, as is witnessed by the value $S_{A} = \log(2)$ in Fig.~\ref{fig:fig1}c (orange dashed line). The opposing limit, where $\varphi=0$ and $\langle H_{\rm M}\rangle/N = -1$, can be regarded as an anneal between $-H_{\rm M}$ and $-H_{\rm P}$. \change{Here the ground state of $-H_{\rm P}$ given by $|{\rm GS}\rangle = \lvert\uparrow\rangle^{\otimes N}$, a product state with energy $E_{\rm GS} = -2N + 1$~\footnote{Notice that in absence of the longitudinal field $\sum_i \sigma_i^z$ the two states $|\uparrow\rangle^{\otimes N}$ and $|\downarrow\rangle^{\otimes N}$ are degenerate with energy $E = -(N-1)$, thus the ground state of $-H_{\rm P}$ corresponds to their even and odd superpositions. However, the inclusion of the longitudinal term shifts $|\uparrow\rangle^{\otimes N}$ down in energy to $E = -2N + 1$ and $|\downarrow\rangle^{\otimes N}$ up in energy to $E = 1$ effectively lifting the degeneracy, leaving $|\uparrow\rangle^{\otimes N}$ as the unique ground state.}, is reached at the end of the ramp, correspondingly $S_A=0$ as also seen in the Figure.} 



For initial states away from the two extremal cases, we find that the evolution reaches final configurations with $S_{A}$ rapidly converging to the Page value of random pure states (colored dashed lines in Fig.~\ref{fig:fig1}c). Interestingly, states with $S_{A}$ closer to the Page value also have mean-energy $\langle H_{\rm M}\rangle \approx 0$, corresponding with the middle of the energy spectrum, an observation which will be further discussed in Sec.~\ref{sec:chaos_in_unitary}. Finally, since $\langle\theta,\varphi|H_{\rm M}|\theta,\varphi\rangle = N\sin(\theta)\cos(\varphi)$, all other $|\theta,\varphi\rangle \notin \mathcal{S}$ have a mean-energy which rapidly decreases away from $\theta = \pi/2$. Hence the fate of the majority of states, $|\theta,\varphi\rangle$, in the manifold of SCSs, with the exception of small neighborhoods in the vicinity of $(\theta,\varphi) = (\pi/2,0), (\pi/2, \pi)$, i.e., $\langle H_{\rm M}\rangle/N = \pm1$, is to be transported by the forward ramp into highly entangled states, displaying an entanglement entropy which grows with system size, and with some of them resembling random pure states.

\subsection{Cyclic ramps}
\label{subsec:scs_cyclic}
We now consider the cyclic ramp. For a perfect adiabatic connection, the two extremal states, $|\pi/2, 0\rangle = |+\rangle^{\otimes N}$ and $|\pi/2, \pi\rangle = |-\rangle^{\otimes N}$, must be transported into themselves, i.e., $|\psi(2T)\rangle = |\psi(0)\rangle = |\pi/2, \{0,\pi\}\rangle$. Hence, we are also interested in determining the degree to which such return takes place. To do so, we look at the final state fidelity $f(2T) = |\langle \psi(2T)| \psi(0)\rangle |^2$, which will complement the study of the entanglement entropy $S_{A}$.

Both $S_{A}(2T)$ and $f(2T)$ for all $|\psi(0)\in\mathcal{S}$ are shown, as function of the mean-energy density $\langle H_{\rm M}\rangle/N$ of the initial state, in Fig.~\ref{fig:figure_scs}a and \ref{fig:figure_scs}b, respectively. We find that the two extremal states of $H_{\rm M}$ are successfully connected back to themselves through the cyclic evolution, as evidence by the corresponding values of $S_{A}(2T) = 0$ and $f(2T) = 1$. On the other hand, most of the initial states reach ``warmer'' states, i.e., $S_{A}(2T)>0$. In particular, the value of $S_{A}(2T)=\log(2)$ (grey dotted line in Fig.~\ref{fig:figure_scs}a) seems to provide a good discriminator between initial states reaching $f(2T)>0$ and $f(2T)\sim0$ when $S_{A}(2T)<\log(2)$ or $S_{A}(2T)>\log(2)$, respectively. Furthermore, most of the energy range defined by $H_{\rm M}$ is occupied by states which fail to come back to themselves with a finite $f(2T)$. In connection with the above discriminator, we see that states with $S_{A}(2T)<\log(2)$ have $|\langle H_{\rm M}\rangle/N| \gtrapprox 0.9$ and those with $S_{A}(2T)>\log(2)$ have $|\langle H_{\rm M}\rangle/N| \lessapprox 0.9$. Notice that the effect is also asymmetric: It is easier to transport along a cyclic adiabatic ramp product states with $\langle H_{\rm M}\rangle$ close to the top of the spectrum than those with $\langle H_{\rm M}\rangle$ close to the bottom of the spectrum of $H_{\rm M}$.

Finally, we stress that the results discussed in Sec.~\ref{subsec:scs_forward} and Sec.~\ref{subsec:scs_cyclic} are not unique to product states; in fact, entangled initial states display a similar behavior. In App.~\ref{app:evol_dicke} we report results for a set of initial states given by the eigenstates of $H_{\rm M}$, i.e., Dicke states, which are in general entangled. This suggests an initial-state-independent character of the reported behavior, connected to universal properties of the unitary operator $U(T_{\rm tot})$. We explore this aspect of the problem in the following section.

\begin{figure*}
\centering
\includegraphics[width=0.98\linewidth]{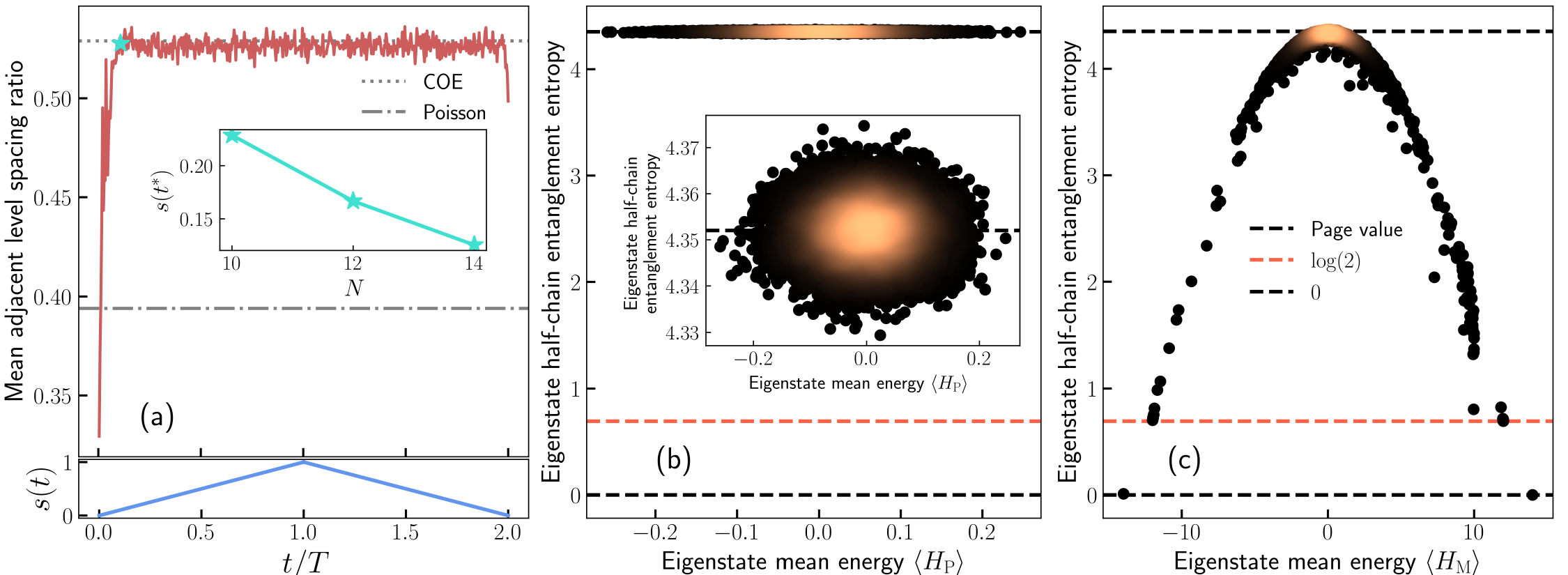}
\caption{Spectral properties of the adiabatic unitary operator for both forward and cyclic ramps. (a) Mean level spacing ratio (MLSR) as a function of time for $0\leq t \leq 2 T$ covering both forward and backward ramps. The green star indicates the first time instant $t^\ast$ at which the unitary becomes fully chaotic ($s\simeq0.14$ for $N=14$), and the inset shows the behavior of this quantity with system size. The dash-dot line shows the MLSR for an integrable system and the dotted line that of a chaotic system. The bottom panels shows $s(t)$, with the forward ramp stopping at $t/T = 1$ and the cyclic ramp defined over the full range of $t/T$. (b) Half-chain entanglement entropy of all the eigenstates of $U(T)$ as a function of their mean energy with respect to the problem Hamiltonian $H_{\rm P}$. (c) Half-chain entanglement entropy of all the eigenstates of $U(2T)$ as a function of their mean energy with respect to mixer Hamiltonian $H_{\rm M}$. Both in (b) and (c) the dots are colored based on the local density, with black indicating very low density and light orange indicating very high density, the dashed lines (bottom to top), indicate $\mathcal{S} = 0, \log(2), 0.5\log(N) - 0.5$, respectively \change{(see App. \ref{app:evol_dicke} for further information about the relevance of these values)}. all panels the parameters are: $N=14$, $T=600$, $dt = 0.5$.}
\label{fig:figure3}
\end{figure*}

\section{Chaos in the unitary evolution operator}
\label{sec:chaos_in_unitary}
In this section we characterize the global spectral properties of the unitary evolution operator describing the adiabatic driving of the Hamiltonian in Eq.~(\ref{eq:hami_s}), as approximated by Eq.~(\ref{eq:approx_adiabatic_unitary}). We investigate the onset of quantum chaos on the annealing process from properties of the eigenphases and eigenvectors of $U(t)$.

The degree of correlation among eigenphases is studied with the MLSR as introduced in Eq.~(\ref{eq:gaps_def}) and Eq.~(\ref{eq:ratios_def}), allowing us to discern whether the unitary can be associated with an integrable, quasi-integrable, or quantum chaotic system. As we deal with spectral properties of a unitary operator, the random matrix ensembles linked with quantum chaotic behavior are now the circular ensembles \cite{haake1991}. In particular, for the circular orthogonal ensemble (COE), $\overline{r}_{\rm COE}\simeq 0.529$. In contrast with the case of Hamiltonian spectra, unitary operators do not have a clear-cut notion of low-energy and bulk sectors in their spectra. As such, we perform the MLSR analysis including \textit{all} the eigenphases of the positive parity symmetry sector.

To study the structure of the eigenstates, we use the half-partition entanglement entropy $S_{A}$ as defined in Eq.~(\ref{eq:half_chain_ee}). For many-body systems in one dimension the half-chain entanglement entropy of individual eigenstates as a function of system size displays two dominant behaviors: i) Area-law entanglement with $S_{A}(N) \simeq K$ with $K$ a constant independent of the system size, and ii) volume-law entanglement with $S_{A}(N) = c_1 N + c_2 + O(N^{-1})$ a linear dependence on the system size. The coefficients $c_{1,2}$ allow us to distinguish between random gaussian states~\cite{Magan2016,Bianchi2021} and random pure states which saturate the Page value~\cite{Page1993}, with $c_1=\log(2)/2$ and $c_2 = -1/2$. Importantly, chaotic eigenstates of many-body Hamiltonians obey a volume-law of the half-chain entanglement entropy~\cite{Amico2008,Bianchi2022,Kliczkowski2023}, thus one can diagnose the presence of this type of eigenstates based on the behavior of $S_{A}$.

\subsection{Forward ramps}
\label{subsec:forward_ramp}
The evolution along the forward ramp is defined iterativelly via a product of unitaries in time succession, see Eq.~(\ref{eq:approx_adiabatic_unitary}). For $t=m \Delta t \in[0, T]$, we write $U(t = m \Delta t) = \left[ \prod_{p=0}^{m-1} e^{-i p\Delta t H(s(p \Delta t))} \right] e^{-i m\Delta t H(s(m \Delta t))}$ where the product inside the square brackets is the unitary up to the preceding time $U((m-1)\Delta t)$. The unitary of the full ramp is obtained when the number of steps equals $T/\Delta t$. The spectral decomposition of any of the unitaries along the ramp reads $U(t)|\mu_l(t)\rangle = e^{-i\mu_{l}(t)}|\mu_l(t)\rangle$, with $\{\mu_l(t)\}_{l=1,..,2^N}$ the set of eigenphases and $\{|\mu_l(t)\rangle\}_{l=1,..,2^N}$ the set of associated eigenvectors, and with $t\rightarrow T$ for the unitary of the full evolution $U(T)$.

We consider first the MLSR of the set of eigenphases $\{\mu_l(t)\}$ of each of the unitaries $U(t)$ for $0\leq t\leq T$, i.e., along the ramp. The results are shown in Fig.~\ref{fig:figure3}a for a system with $N=14$ spins. Naturally, at $s(t\simeq 0)$ the MLSR indicates the global properties of the unitary coincide with those of an integrable system, which is expected given that $H(s(0))=H_{\rm M}$. Nonetheless, we find that once the ramp $s(t)$ reaches a value as small as $s(t^\ast)\simeq0.14$ (green star in in Fig.~\ref{fig:figure3}a), the unitary already shows a MLSR agreeing with that of a COE random matrix, indicating the evolution has become quantum chaotic. Note that at that point in the interpolation path ($s\simeq 0.14$), the instantaneous Hamiltonians $H(s)$ are still far from the fully chaotic regime, as can be seen from Fig.~\ref{fig:figure2}a. Interestingly, analyzing $s(t^\ast)$ vs $N$ shows a clear decreasing tendency with increasing $N$ (see inset in Fig.~\ref{fig:figure3}a). Thus, we observe that even the `gentle' adiabatic driving has a notable effect by leading to an early emergence of ergodic properties in the unitary evolution operator.  Further, we observe that for the rest of the forward ramp, \textit{i.e.}, $s(t^\ast)< s(t) < 1$, the adiabatic evolution unitary remains quantum chaotic. 

Having observed the emergence of chaotic properties along the forward ramp as witnessed by the properties of the eigenphases, we now focus on the unitary of the complete forward ramp, $U(T)$, and its associated set of eigenstates $\{|\mu_l(T)\rangle\}_{l=1,..,2^N}$. In Fig.~\ref{fig:figure3}b we plot the half-chain entanglement entropy of all eigenstates of $U(T)$ as a function of their mean energy with respect to $H_{\rm P}$, the Hamiltonian at the end of the forward ramp. As anticipated by the MLSR analysis, the forward ramp leads to spectral properties of the unitary coinciding with those of a COE random matrix, thus most eigenstates have a $S_A$ which is consistent with the Page value for random pure states. We emphasize that there is no dependence on the final energy, i.e. we observe no signatures of the presence of a low energy or high energy sectors. The inset shows a closer view of the values of $S_{A}$ where a Gaussian distribution centered at $(\langle H_{\rm P}\rangle, S_{A}) = (0, 7\log(2)-1/2)$ is apparent. Thus, our analysis shows that the dynamical map $U(T)$ describing the adiabatic evolution is highly chaotic, and spectrally indistinguishable from a random quantum circuit.

\subsection{Cyclic ramps}
\label{subsec:cyclic_ramp}
The onset of quantum chaos becomes more subtle (and thus, more interesting) when we consider the evolution along the cyclic ramp. Since the cyclic ramp is composed of a forward and a backward ramps, the first half of the evolution is always described by $U(T)$. Hence, the results of the MLSR study discussed in Sec.~\ref{subsec:forward_ramp} apply here for the first half of the evolution. For the second half of the evolution completing the cyclic ramp, the results of the MLSR analysis are shown in the same Fig.~\ref{fig:figure3}a, for the regime $T\leq t\leq 2T$. We notice that for most of the backward ramp the chaotic properties remain and only for the final portion, $s(t)\gtrapprox1.9$, we observe a deviation from the COE value, with the MLSR of the eigenphases of $U(2T)$ acquiring an intermediate value between $\overline{r}_{\rm Int}$ and $\overline{r}_{\rm COE}$. Thus, while there is still some degree of level repulsion, the system no longer resembles a fully random unitary. \change{In fact, there exists a choice of Hamiltonian with respect to which the mean energies of eigenstates together with their entanglement entropies $S_{A}$ allows us to consistently divide the spectrum into sectors of low- and high-energy, and bulk. For us the choice of reference Hamiltonian is informed by the structure of the cyclic ramp. This means that the spectral structure of $U(2T)$ is more similar to that a chaotic Hamiltonian system, than of a random unitary. We devote the rest of this section to a judicious quantitative study supporting our previous claim.}

\change{The entanglement entropy $S_{A}$ of eigenstates allows us to distinguish ergodic ($S_{A}$ saturating the Page value) from nonergodic ($S_{A}$ constant) eigenstates. Generic many-body Hamiltonians have eigenstates of both classes. For integrable and near-integrable Hamiltonians, ergodic eigenstates, if any, are found at energies close to the middle of the spectrum, while nonergodic eigenstates may be found at all energies~\cite{Beugeling2015}. For a nonintegrable chaotic Hamiltonian the two types of eigenstates occupy distinct energy ranges. In fact, as $E\to\overline{E}$ with $\overline{E} = {\rm tr}(H)/2^N$ the mean energy, from above or below, a crossover between $S_A\propto N^0$ and $S_A\propto N^1$ takes place~\cite{Miao2021}, indicating that nonergodic (area-law) eigenstates only exist at the edges of the spectrum (with some notable exceptions~\cite{serbyn2021}). As such, $S_A$ as a function of $E$ for the eigenstate of many-body Hamiltonians is a concave function of the energies with a maximum at the ``middle'', i.e., the mean energy $\overline{E}$, with a characteristic ``arch'' shape. Importantly, for chaotic Hamiltonians the average and variance of $S_A$ on only the mid-spectrum eigenstates follows well characterized behaviors~\cite{Garrison2018,Haque2022,Kliczkowski2023}.}

\begin{figure}
\centering
\includegraphics[width=0.97\columnwidth]{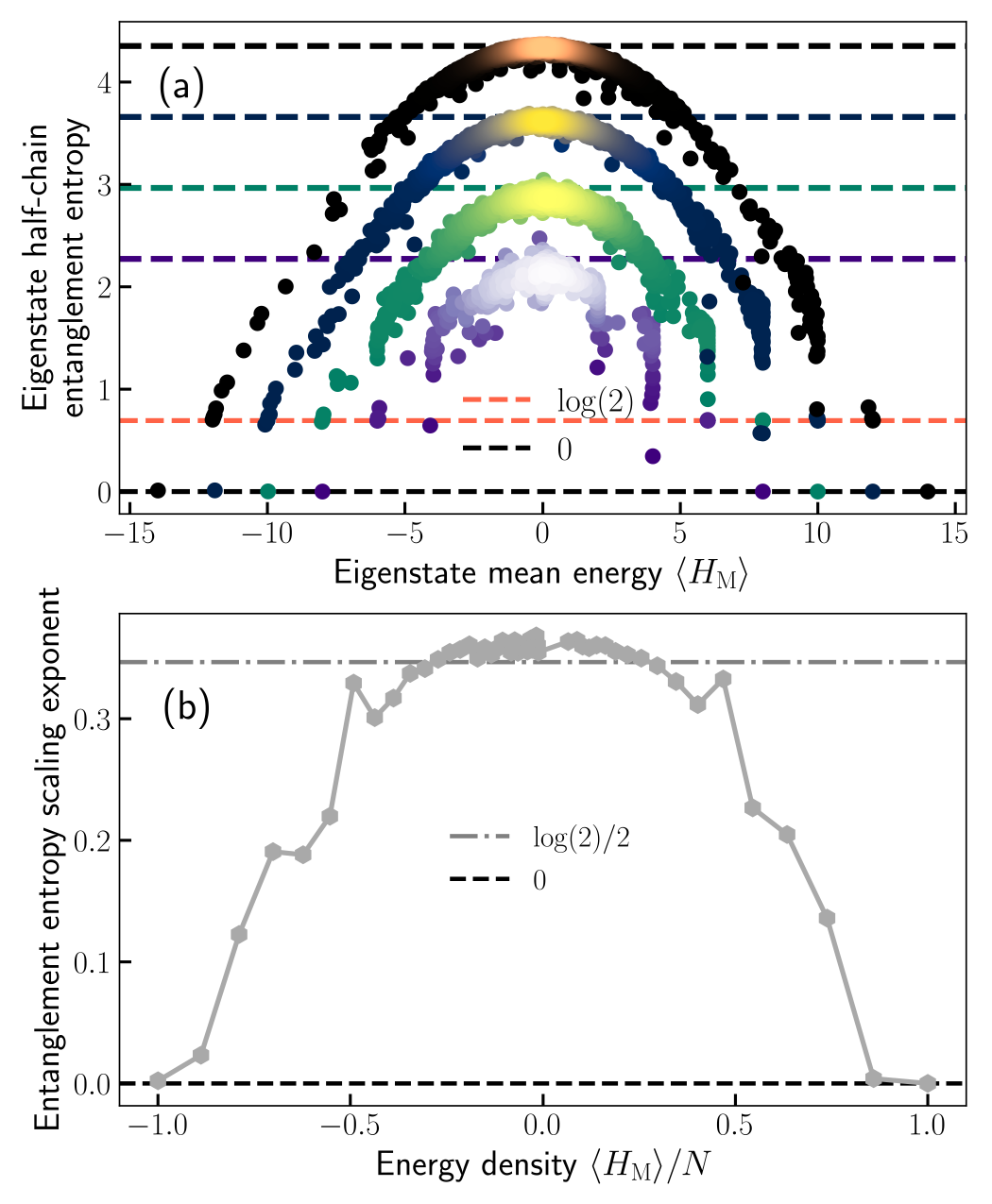}
\caption{(a) Half-chain entanglement entropy of all the eigenstates of $U(2T)$ as function of their mean energy with respect to $H_{\rm P}$. Darker color indicates low point density, lighter color indicate high point density. Purples, greens, blues, black, correspond to system sizes of $N = 8,10,12,14$, respectively. The bottom  two dashed lines indicate $\mathcal{S} = 0, \log(2)$, the top four dashed lines indicate the Page value of the entanglement entropy for the respective system size. (b) Scaling exponent with system size of the locally averaged half-chain entanglement entropy. A rapid convergence to the Page value of $\log(2)/2$ indicates that most of the eigenstates considered in (a) exhibit volume-law entanglement entropy and thus are ergodic states.}
\label{fig:figure4}
\end{figure}

We thus focus on $U(2T)$ and ask: how similar is the structure of its egenstates $\{|\mu_l(2T)\rangle\}_l$ as seen in the plane $(\langle H_{\rm M}\rangle, S_{A})$ to that of a chaotic Hamiltonian? Here $H_{\rm M} = H(0) = H(2T)$ defines the natural energy scale. In Fig.~\ref{fig:figure3}c we show $S_{A}$ vs $\langle H_{\rm M}\rangle$ for all the $\{|\mu_l(2T)\rangle\}$ for a system of $N=14$ spins. Clearly the data follows a curve with all the characteristics of the $S_A$ vs $E$ curve of a chaotic Hamiltonian described before. In particular, we observe a concave shape with a maximum at the mean energy; we provide a semi analytical form for this function in App.~\ref{app:ee_eigenstates}, together with an analysis of the expected behavior of the mean and variance of $S_A$ for the ``mid-spectrum'' eigenstates (see App.~\ref{app:ee_eigenstates} for details). Further, the structure seen in Fig.~\ref{fig:figure3}c shows clear emergent energy sectors which can be classified into low-energy, bulk, and high-energy, distinctions which are absent in the forward-ramp unitary.

\newchange{Physically, then, we observe sharp difference between evolutions stemming from the the forward and cyclic annealing schedule. While the forward dynamics is described by a unitary process which is fully chaotic across all energy scales, the cyclic dynamics shows an emergent structure in which low- and high-energy eigenstates states deviate substantially from the typical chaotic predictions. As we will confirm in the next section, this deviation can be linked to adiabaticity in those parts of the energy spectrum}. 

\newchange{We also note that the} lack of adiabaticity in the bulk of the spectrum is evident in the preceding analysis. A system evolving in perfect adiabatic fashion should connect every eigenstate of $H_{\rm M}$ to itself after the cyclic ramp. Thus, in the $S_A$ vs energy diagram, one would observe that only discrete values of $\langle H_{\rm M}\rangle$ are present (in this case, this would be $N+1$ distinct eigenvalues). In contrast to this, we find the values of $\langle H_{\rm M}\rangle$ appear to be continuous in value, particularly near $\langle H_{\rm M}\rangle\approx0$.

To complete our investigation of the structure of eigenstates of $U(2T)$, we look into the behavior of $S_{A}$ as a function of the system size $N$. In Fig.~\ref{fig:figure4}a we show $S_{A}$ vs $\langle H_{\rm M}\rangle$ for all the eigenstates of $U(2T)$ for systems with $N=8,10,12,14$ (purples, greens, blues, black, respectively). We find that, even for the smallest size considered, signatures of a structure resembling that of a nonintegrable Hamiltonian are noticeable. We thus study $S_{A}(N)$ in terms of an energy density $E_{\rm d} = \langle H_{\rm M}\rangle/N \in[-1, 1]$. We bin the $E_{\rm d}$ axis into small energy-density windows $E_{\rm W} = [E_{\rm d} - \Delta E_{\rm d}, E_{\rm d} + \Delta E_{\rm d}]$, whose size is chosen as to not have empty windows. For each system size we count the number of eigenstates inside a given energy-density window $N_{\rm W} = |\{S_{A}(\epsilon) : \epsilon\in E_{\rm W}\}|$ with $|.|$ indicating the cardinality of the set, and compute the energy-window averaged $S_{A}$, that is $\left.\overline{S_{A}}\right|_{E_{\rm W}} = \frac{1}{N_{W}}\sum_{\epsilon\in E_{\rm W}}S_{A}(\epsilon)$. Finally we regress the energy-density window averaged $S_{A}$'s to a linear model, that is,  $\left.\overline{S_{A}}\right|_{E_{\rm W}}(N) \leftarrow aN + b$ with $(a,b)$ the regression coefficients. 

In Fig.~\ref{fig:figure4}b we show the values of $a$ as a function of $E_{\rm d}$. We recognize two distinct behaviors, volume-law entanglement and area-law entanglement, with the particularity that the volume-law eigenstates can be subdivided into two different groups. For $|E_{\rm d}|<0.5$ we can unambiguously conclude that the eigenstates of $U(2T)$ obey a volume-law of entanglement which is consistent with the Page result for random pure states, and that the majority of eigenstates are found in this region. For $0.5 < |E_{\rm d}| \lessapprox 0.8$, we encounter few sets of eigenstates which do follow a volume-law of entanglement but whose values of $(a,b)$ indicate they might be random Gaussian states~\cite{Bianchi2021}. Finally, for $|E_{\rm d}|>0.8$ we observe four sets of eigenstates which obey an area-law of entanglement. Interestingly for all system sizes studied, we can readily recognize the states in this last category. Naturally, the lowest and highest energy states, at $E_{\rm d} = \pm1$, correspond to the ground state  $|-\rangle^{\otimes N}$ and highest excited state $|+\rangle^{\otimes N}$ of $H_{\rm M}$, which appear both in Fig.~\ref{fig:figure3}c and Fig.~\ref{fig:figure4}a at $S_{A} = 0$ (first black dashed line). The next one, in increasing $S_{A}$, corresponds to Dicke states with one flip (low-energy) and $N-1$ flips (high-energy), and states with similar entanglement structure, appearing at $S_{A}=\log(2)$ (orange dashed line) in Fig.~\ref{fig:figure3}c and Fig.~\ref{fig:figure4}a.

\section{Operator dynamics and information scrambling}
\label{sec:scrambling}

In this section we analyze the quantum annealing dynamics by focusing on the Heisenberg evolution of different classes of operators. In particular, we will study the size of operators as a function of time, which is a standard metric in the study of quantum information scrambling \cite{Roberts2018,Parker2019,zhuang2019,omanakuttan2023}. During scrambling dynamics, initially local operators expand towards the rest of the degrees of freedom of the system, in such a way that their support becomes increasingly nonlocal and the so-called mean operator size $\mu(t)$ (to be defined below) becomes extensive. For general unitary dynamics, this process can be described by the operator size distribution (OSD) $\{P_k(t)\}$, $k=1,\ldots, N$, where $P_k(t)\geq 0$ and $\sum_k P_k(t)=1$. For a given initial operator $A(0)$, $P_k(t)$ quantifies the support of the Heisenberg-evolved operator $A(t)=U(t)^\dagger A(0) U(t)$ in the set of Pauli operators $\{Q\}_{r(Q)=k}$ of size exactly $k$. Here the size $r(Q)$ of a multi-body Pauli operator equates the Hamming weight of the associated Pauli string.  Formally, the OSD is defined as
\begin{equation}
    P_k(t) = \frac{1}{d^2\: \trace{A^2}}\sum\limits_{r(Q)=k} \lvert \trace{Q\:U(t)^\dagger A U(t)}\rvert^2,
\end{equation}
where $d=2^N$. In turn, the mean operator size is defined as the first moment of this distribution,
\begin{equation}
    \mu(t) = \sum\limits_{k=1}^N k P_k(t).
    \label{eq:mean_op_size}
\end{equation}

Evidently, $1\leq \mu(t)\leq N$. For local operators, one expects $\mu(t)$ to grow at short times (in some cases, exponentially \cite{schuster2023}) and, under generic dynamics, the mean operator size becomes extensive, i.e., proportional to the system size $N$. Particularly, a Haar-random unitary leads to $\mu_{\rm Haar} \simeq 3N/4$ independent of our choice of initial Pauli operator. 
Naturally, higher order moments of the OSD can also be considered as more refined descriptions of the scrambling process \cite{Roberts2018,omanakuttan2023}. 

We will consider the dynamics of two types of initial operators: single-site and global. Single-site operators correspond to size-one Paulis, e.g.
\begin{equation}
    A(0) \rightarrow \{\sigma_{i_0}^x,\sigma_{i_0}^y,\sigma_{i_0}^z\}
    \label{eq:local_ops}
\end{equation}
where we take $i_0=\floor{N/2}$ as a generic choice. The set of global operators is formed by the collective spin components
\begin{equation}
    A(0) \rightarrow \{S_x,S_y,S_z\}
    \label{eq:global_ops}
\end{equation}
where $S_\alpha = \sum\limits_{i=1}^N \sigma_i^\alpha/2$. Clearly, for these choices of single-site and global operators we have $\mu(0)=1$ as these operators have (initially) only support on size-1 Paulis. Thus, we are interested in analyzing the initial growth and subsequent evolution of the mean operator size under the different types of adiabatic evolution considered in the previous section. 

\begin{figure}
\centering
\includegraphics[width=\columnwidth]{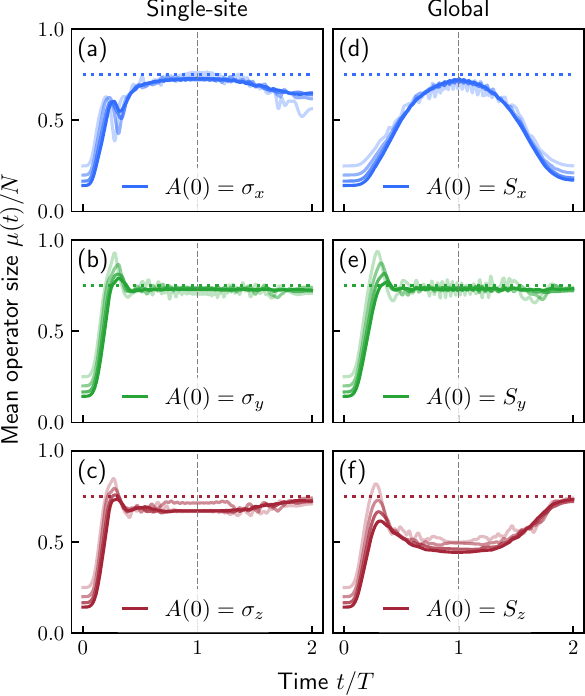}
\caption{Normalized mean operator size evolution $\mu(t)/N$, defined in Eq.~(\ref{eq:mean_op_size}), for different choices of initial operators. Plots (a)-(c) correspond to initial single-site operators shown in Eq.~(\ref{eq:local_ops}), while (d)-(f) to the global operators defined in Eq.~(\ref{eq:global_ops}). Data is shown for system sizes $N=4,5,6,7$ (light to dark color tones). All cases show evolution under the cyclic driving discussed in Sec.~\ref{sec:model}; dashed vertical lines indicate $t=T$, where the forward ramp ends, and the backward ramp begins. Dotted horizontal lines correspond to the mean operator size expected for Haar random evolution $\mu_{\rm Haar}/N=3/4$.}
\label{fig:figure5}
\end{figure}

\subsection{Dynamics of local operators}

Fig.~\ref{fig:figure5} (a)-(c) shows the evolution of the mean operator size for the set of single-site operators described above. Data is shown for the complete cyclic ramp, meaning that dynamics up to $t/T=1$ corresponds to the forward ramp from $s(0)=0$ to $s(T)=1$, while the subsequent evolution arises from the backwards ramp leading back to $s(2T)=0$. In all cases, we show data for different system sizes $N=4,5,6,7$ (light to dark tones). We point out that exact numerical evaluation of the OSD is computationally expensive as it requires to calculate the full operator evolution and then project onto each element of the many-body Pauli basis, which is exponential in size.
The evolution shown in Fig.~\ref{fig:figure5} (a)-(c) indicates that single-site operators grow during the forward adiabatic evolution and become extensive, with support on all the degrees of freedom of the system. Notably, the operator size reached by the end of the forward ramp at $t=T$ is very close to that expected for random evolution, $\sim 3N/4$. Thus, in line with the results in Sec.~\ref{sec:evol_scs} and Sec.~\ref{sec:chaos_in_unitary}, we see that the adiabatic driving leads to a highly chaotic evolution, in this case witnessed by scrambling in the operator dynamics. 

Perhaps more surprisingly, we observe that the operators continue to be extensively delocalized even for $T\leq t \leq 2T$, that is, during the backward ramp. The most striking case is $\sigma_y$, where the operator size equilibrates at the value expected for random evolution and shows no sign of reversing (``shrinking'') to lower values at the end of the protocol. The evolution of these single-site operators thus starkly contrasts the cyclic evolution expected for states in the low-energy sector. In fact, we find that single-site operators evolve in an \textit{almost} thermal fashion, with effectively irreversible dynamics. This hints at the fact that the evolution of these operators is dominated by the properties of the bulk eigenstates of $H(s(t))$, where adiabaticity is essentially not possible. 

\subsection{Dynamics of global operators}

As we discussed above, the operator dynamics for the set of single-site Pauli operators shows clear signatures of information scrambling. In particular, we found that the mean operator size dynamics is equivalent to that generated by Haar-random dynamics, meaning that the ``gentle'' adiabatic drive has a similar effect as one would expect from a thermalizing periodically driven system, or even from a random quantum circuit. \change{However, as we have seen in Sec.~\ref{sec:chaos_in_unitary}, the evolution stemming from the cyclic path showed clear deviations from chaotic behavior. Here we investigate such behavior for the case of operator dynamics.}


In Fig.~\ref{fig:figure5} (d)-(f) we display the mean operator size for the global operators $\{S_x,S_y,S_z\}$. The dynamics bears similarities and differences with the evolution of single-site operators. For the forward ramp, $0\leq t \leq T$, we observe that operator sizes become clearly extensive with values matching those of random operators for $S_x$ and $S_y$; the case of $S_z$ shows maximum operator size increasing with system size, but clearly deviates from the Haar prediction. \change{We attribute this to the fact that the operator commutes with the interacting Hamiltonian $H_P$ (notice that a similar, but weaker effect, is observed for the local Pauli $\sigma_z$), which is responsible for the scrambling behavior}.

From all cases considered, the one that stands out the most is $S_x$, where we observe that size of $S_x$ shrinks back to $\sim 1$ during the backward ramp. Thus, we see that the dynamics of $S_x$ reverses back to its initial configuration, although not perfectly. A key aspect here is that $S_x$ is a special operator in this problem, since its proportional to the mixer Hamiltonian, i.e. $H(0)=H(2T)=H_M=2 S_x$. At first sight, the evolution of $S_x$ seen here is consistent with the expectation that the adiabatic evolution should connect the eigenstates of $H_M$ back to themselves at the end of the cyclic protocol. However, a more detailed inspection reveals that only the eigenstates of $S_x$ near the edges of its spectrum follow this expected evolution, while eigenstates near the bulk undergo mixing evolution, \change{due to imperfect adiabaticity in that regime}. We show this analysis in Appendix~\ref{app:evol_dicke}. As evidenced from the results in this section, it suffices that the dominant eigenstates (i.e. those with larger eigenvalues in absolute value) evolve adiabatically for the operator size to shrink back to its original configuration.
\change{Thus, in line with the results obtained in Sec. \ref{subsec:cyclic_ramp}, we find a non-trivial balance between strong, random-like chaotic behavior (e.g. as seen in the the behavior of $S_y$), and clear signatures of the near-adiabatic evolution leading to non-generic dynamics (e.g. the operator shrinking of $S_x$), including cases exhibiting a mixture of both phenomena (like $S_z$).}

\subsection{Connection to out-of-time ordered correlators and route to experimental exploration}

\change{Witnessing chaotic signatures in quantum many-body systems is a challenging task for experiments. This is due to the fact that many measures of chaoticity are computed over many-body eigenstates, which are typically hard to prepare. Even for those states which can be accessed (for instance following a quench), typical local correlation functions struggle to witness chaotic behavior, and more intricate measures are needed. This is evidenced from our own analysis up to here, where we needed to compute quantities like entanglement entropies, unitary eigenphases, or operator wavefunctions. 

Nonetheless, significant attention has been given to so-called out-of-time-ordered correlators (OTOCs) as a way to probe chaotic behavior in experiments. These quantities take the form 
\begin{equation}
    C_{A,B}(t) = \trace{A(t)^\dagger B^\dagger A(t) B}
    \label{eq:otoc}
\end{equation}
where $A$ and $B$ are general operators and $A(t)=U(t)^\dagger A U(t)$. Many proposals have been put forward in the past decade to measure OTOCs leveraging different experimental capabilities \cite{swingle2016,garttner2017,vermersch2019,landsman2019}, and recently many of them have been successfully demonstrated in, e.g., superconducting qubits \cite{scr_scqubits2021}, Rydberg atoms \cite{scr_ions2020}, and trapped ions \cite{scr_rydberg2024}. Importantly for our discussion, OTOCs have a direct connection to the mean operator size $\mu(t)$; as this can be recovered by considering the operator $B$ in Eq. (\ref{eq:otoc}) to be single-site Paulis. More specifically, one has that \cite{Roberts2018,omanakuttan2023}
\begin{equation}
    \mu(t) = \frac{3N}{4}\left(1-\mathcal{M}(t)\right),
\end{equation}
where
\begin{equation}
    \mathcal{M}(t) = \frac{1}{3N}\sum\limits_{r(B)=1} C_{A,B}(t)
\end{equation}
i.e. an average over all OTOCs with single-site Paulis. Thus, we see that the scrambling behavior analyzed in this section has a route for potential experimental exploration, which is rooted in the extensive studies of OTOCs in various platforms. }

\section{Discussion and outlook}
\label{sec:discussion_outlook}

In this work we have shown that a basic model of a quantum annealer can display clear signatures of quantum ergodicity and scrambling. These features become prominent when one considers the dynamical behavior of these models beyond the low (and high) energy regime. Here, we have analyzed three specific aspects which fall into this criterion: the generation of volume-law entanglement when the system is initialized in a product state which is rotated from the ground state, the random-matrix-like spectral properties of the complete unitary map $U(T)$ describing the annealing dynamics, and the onset of information scrambling as measured by the growth of generic operators in the Heisenberg picture. Furthermore, we have shown that the counterintuitive fact that slow adiabatic drive results in strongly chaotic dynamics is balanced by the existence of clear deviations from chaoticity which are observed when a cyclic adiabatic drive is considered. \newchange{In this regime, we found that adiabaticity plays a major role in explaining such deviations. A key example of this is the observation of reverse scrambling in the Heisenberg dynamics of certain collective operators}.

Our findings show that a quantum annealing device, originally designed to follow adiabatically low-energy states, could have interesting additional applications if one uses it in a different regime. For instance, achieving scrambling, random-like dynamics can be a useful tool for device benchmarking. In fact, while it is somewhat straightforward to engineer a quantum circuit that scrambles most initial states, it is not so clear how to find a class of models where a subset of initial states thermalize while others do not. A notable example of such models would be those featuring Quantum Many-Body Scars (QMBS) \cite{serbyn2021}, which refer to the existence of anomalous high-energy eigenstates which deviate from the eigenstate thermalization hypothesis description~\cite{Srednicki1994,Deutsch1991,Rigol2008}. Systems displaying QMBS can lead to non-thermal dynamics when starting from certain physical states, but evidence suggests that QMBS emerge in fine-tuned models, and are not typically robust (see \cite{lerose2023} for exceptions). Our analysis shows that quantum annealers lead to a dynamical process which will scramble most initial states, except for those associated with the low- and high-energy subspaces of $H_{\mathrm{M}}$.  Having an initial-state dependent entanglement growth could be a useful tool for benchmarking quantum devices, as it would allow  to vary the hardness of tracking the dynamics with classical methods such as matrix product states.

The emergence of ergodicity in the annealing dynamics can also have negative consequences. For instance, the rather sharp separation between low-energy states (following adiabatic dynamics) and bulk states (following mixing dynamics) shown in Fig.~\ref{fig:figure5} indicates that adiabatic algorithms could have limited scope in the task of preparing excited states.

Beyond its relevance for quantum annealing devices, our study is naturally connected to fundamental aspects of nonequilibrium many-body quantum systems under adiabatic driving. \change{A straightforward extension of our work would be to study the dependence of the chaos signatures on the rate of adiabatic driving, i.e. $\dot{s}(t)$. Particularly, a faster sweep of $s(t)$ would bring the dynamics closer to a standard ‘quench’ where the system is changed quickly between two noncommuting (static) Hamiltonians. The deviation from adiabaticity is expected to lead to stronger mixing; however, we have shown here that the adiabatic regime shows features typically connected to driven systems (like the CUE-like spectra). Thus, we believe a nontrivial balance between adiabaticity and chaos would be accessible in such regime. } 

\change{Another interesting path for future exploration is the assessment of how integrability of $H(s)$ would affect the results shown here}. While here we have studied a model defined on a chaotic path (see Fig.~\ref{fig:fig1}a), setting $\lambda = 0$ in Eq.~(\ref{eq:hami_s}) would actually render the model integrable for all values of $s$. \change{In such case, preliminary results indicate that} adiabaticity in the bulk of the spectrum could be easier to achieve thanks to the existence of conserved quantities leading to exact crossings in the spectrum. Recent studies have focused on a complimentary model featuring all-to-all interactions, but where the addition of driving leads to quantum chaos signatures \cite{Puebla2024}. In this vein, we notice that a small $\lambda > 0$ (i.e., a weak integrability breaking) leads to avoided crossings with very small gaps, in fact smaller than in the chaotic regime, which displays level repulsion. The characterization of integrability-to-chaos transition in terms of adiabatic process has also been studied recently \cite{pandey2020,varma2024}, and integrable many-body adiabatic dynamics has been shown to impact the proliferation of errors in quantum annealing \cite{Schiffer2024}. In this context, further characterizing the interplay between adiabaticity and (non) integrability in a more general setting stands out as an interesting open problem.



\acknowledgments
This material is partially based upon work supported by the U.S. Department of Energy, Office of Science, National Quantum Information Science Research Centers, Quantum Systems Accelerator (QSA). Additional support is acknowledged from the Canada First Research Excellence Fund.

\appendix
\section{Evolution of entangled states}
\label{app:evol_dicke}

\begin{figure}[t!]
\centering
\includegraphics[width=0.95\columnwidth]{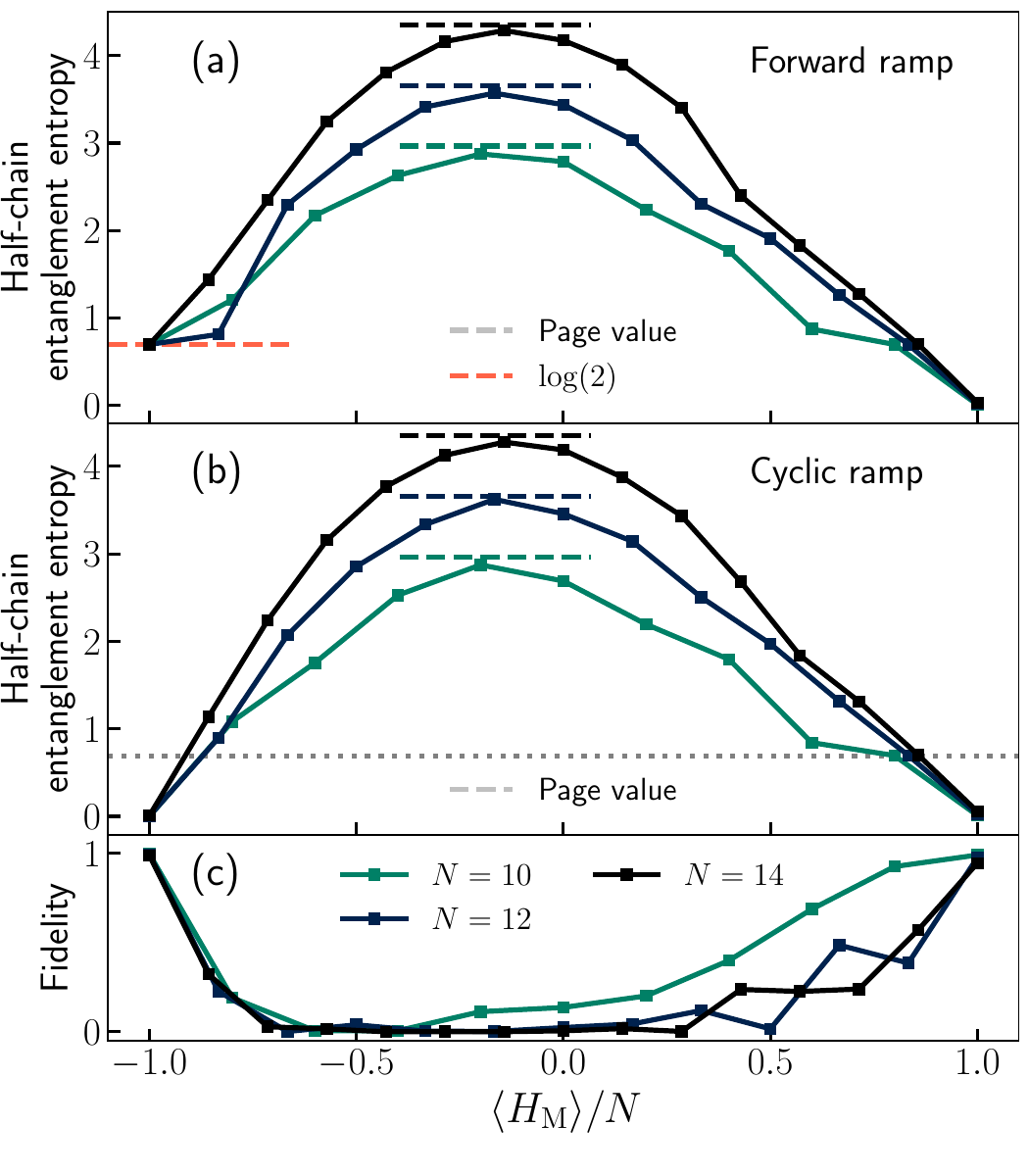}
\caption{Adiabatic evolution of Dicke states in the $x$-basis as a function of their mean energy density $\langle H_{\rm M}\rangle/N$. (a) Half-chain entanglement entropy of the final state after the forward ramp. (b) Half-chain entanglement entropy of the final state after the cyclic ramp. (c) Final state fidelity after the cyclic ramp. Parameters are: $T = 600$, $\Delta t = 0.05$.}
\label{fig:fig_dicke}
\end{figure}
To complement the results on the evolution of product states discussed in Sec.~\ref{sec:evol_scs}, here we consider the evolution of entangled states for both the forward and cyclic ramps. We focus on the family of Dicke states, $\{|D_{k}^N\rangle\}_{k = 0,...,N}$, on the $x$-basis, i.e., the eigenbasis of $H_{\rm M}$. Each of the $|D^N_k\rangle$ is the equal superposition of all $N$-qubit states $|\mathbf{z}\rangle$, where $\mathbf{z}$ is a bitstring of fixed Hamming weight ${\rm w}(\mathbf{z}) = k$, with the latter being the number of spin flips required to map $\mathbf{z}$ on to the all-down (all-zeros) string. That is, a state where $k$ spin flips are distributed permutationally invariant over the $N$ spins. We write
\begin{equation}
\label{eq:dicke_states}
|D_{k}^N\rangle = \binom{N}{k}^{-\frac{1}{2}} \sum_{\mathbf{z}|{\rm w(\mathbf{z}) = k}} \mathrm{H}^{\otimes N} |\mathbf{z}\rangle,
\end{equation}
where $\mathrm{H}$ is the Hadamard quantum logic gate. Clearly $H_{\rm M}|D_k^N\rangle = (-N + k)|D_k^N\rangle$, and the first and last Dicke states correspond to $|D_0^N\rangle = |-\rangle^{\otimes N}$ and $|D_N^N\rangle = |+\rangle^{\otimes N}$ . With the exception of the two extremal states mentioned before and the Dicke states with $k=1,N-1$, all of the other $|D_k^N\rangle$ have a half-chain entanglement entropy which grows logarithmically with $N$, $S_A\sim\log(N)$~\cite{Kumari2022}. In particular, when $k=1$, $|D_1^N\rangle$ is the $N$-qubit W state, which has $S_{A} = \log(2)$. Correspondingly $|D^N_{N-1}\rangle = X^{\otimes N}|D_N^k\rangle$ also has $S_{A} = \log(2)$.

The results of the adiabatic evolution of Dicke states, i.e., with $|\psi(0)\rangle = |D^N_k\rangle$ are shown in Fig.~\ref{fig:fig_dicke} as function of their mean energy density $\langle H_{\rm M}\rangle/N$ for three systems sizes $N=10,12,14$ (green, blue, black, respectively). Let us consider first the forward ramp, Fig.~\ref{fig:fig_dicke}a shows $S_{A}$ of the final state for all the Dicke states. Naturally $|D_0^N\rangle = |-\rangle^{\otimes N}$ is mapped to the ground state of $H_{\rm P}$ which can be evidenced by the value $S_{A}=\log(2)$ (orange dashed line), and the $|D_N^N\rangle = |+\rangle^{\otimes N}$ is mapped to the product state of highest energy of $H_{\rm P}$ with $S_A=0$. Although there are few Dicke states which seem to be transported to states with $S_A(0)=S_A(T)$, see for instance $|D_{N-1}^N\rangle$ for which we have $S_A(0) = S_A(T) = \log(2)$, there is a general tendency of reaching final states with larger $S_A$, with those closer to $\langle H_{\rm M}\rangle/N=0$ reaching $S_A$ consistent with the Page value for random pure states (colored dashed lines). Finally, we notice that the curve $S_A(\langle H_{\rm M}\rangle/N)$ is assymetric, implying that Dicke states closer to the top of the energy spectrum suffer less from the effects of the ergodic properties of the forward ramp unitary.

We move now to discuss the results of the cyclic ramp. Both the final state $S_A$ and fidelity are shown in Fig.~\ref{fig:fig_dicke}b,c, respectively, as function of $\langle H_{\rm M}\rangle/N$. We see that the two extremal states of $H_{\rm M}$ are transported back onto themselves. All the other Dicke states are transported onto states of increasing final $S_A$ with $\langle H_{\rm M}\rangle/N$ reaching values very close to Page value (see dashed lines in Fig.~\ref{fig:fig_dicke}b), thus leading to rapidly decreasing fidelity with $\langle H_{\rm M}\rangle/N$. Interestingly the asymmetry between Dicke states of negative and positive mean-energy is manifested in the final state of the cyclic ramp as well. In particular, those Dicke states of positive mean-energy suffer less from the deleterious effects of the ergodic properties of $U(2T)$ and we observed higher final state fidelities after the cyclic ramp.

\begin{figure}[t!]
\centering
\includegraphics[width=0.95\columnwidth]{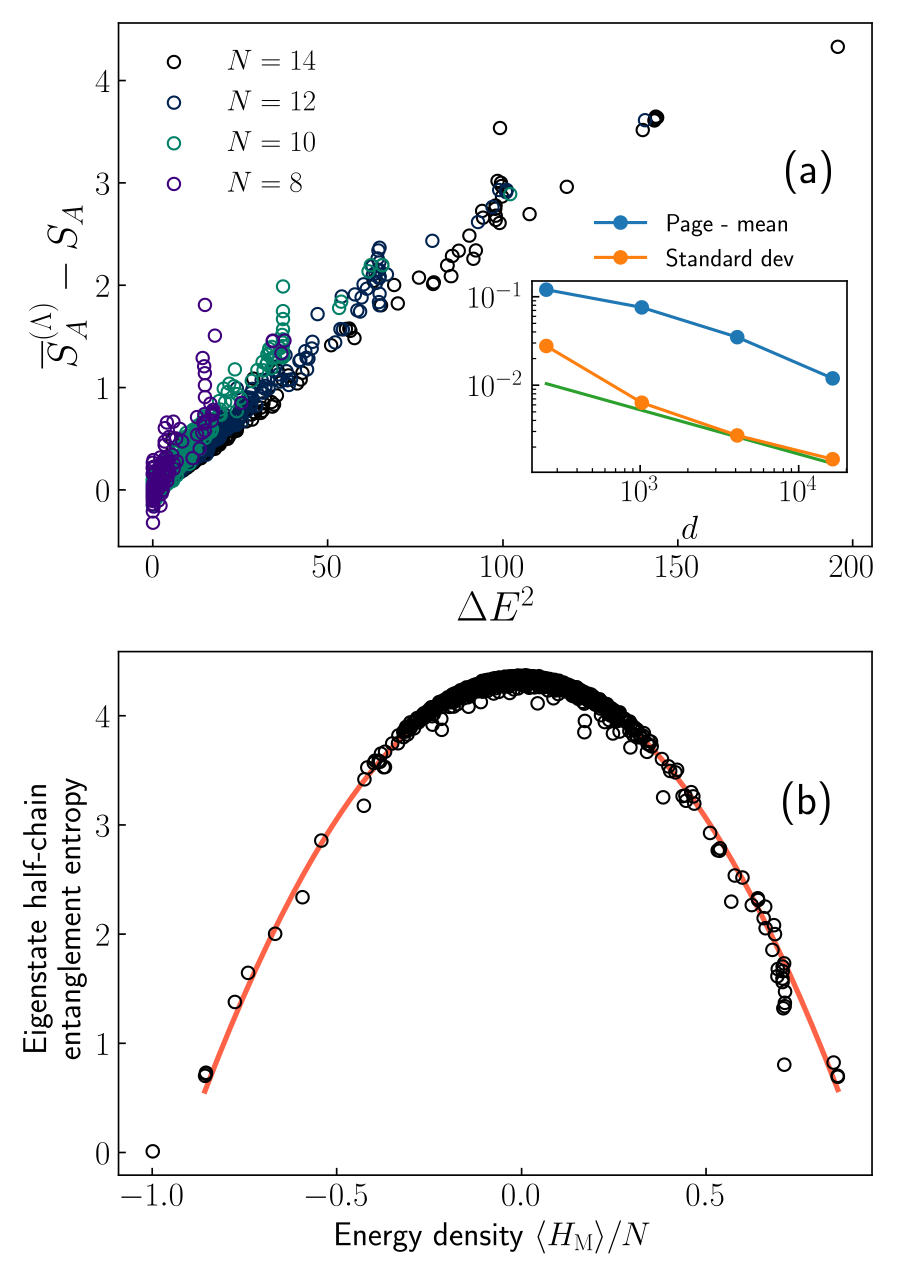}
\caption{\change{Half-chain entanglement entropy of eigenstates of the cyclic ramp unitary $U(2T)$. (a) Deviation from the average $\overline{S}_A^{(\Lambda)}$ as a function of $\Delta E^2$ for the systems sizes, $N = 8,10,12,14$ (purple, green, blue, black), studied in the main text. The inset shows the difference between the Page value of $S_A$ and $\overline{S}_A^{(\Lambda)}$ (blue), and the variance of $S_A$ computed for the $\Lambda$ mid-spectrum eigenstates (orange), both as function of the Hilbert space dimension $d = 2^N$. The green solid line shows the scaling $d^{-1/2}$. (b) $S_A$ vs $\langle H_{\rm M}\rangle$ for the system with $N=14$ (black empty circles), together with the fit obtained from Eq.~(\ref{eq:parabola_fit}) (solid orange line).}}
\label{fig:fig_fitted_parabola}
\end{figure}

\change{
\section{Entanglement entropy of cyclic ramp unitary eigenstates}
\label{app:ee_eigenstates}
In this appendix we give a semianalytical argument for the shape of the $(\langle H_{\rm M}\rangle, \mathcal{S}_A)$ in Fig.~\ref{fig:figure2}c for the cyclic ramp unitary, and share additional results on the behavior of the entanglement entropy for the ``mid-spectrum'' eigenstates. 
}

\change{The bipartite entanglement entropy $S_A$ as function of energy $E$ for the eigenstates of local Hamiltonian is a concave function of the energies with a maximum at the mid energy $\overline{E} = {\rm tr}(H)/2^N$~\cite{Beugeling2015,Garrison2018,Miao2021,Haque2022,Kliczkowski2023}. In particular for nonintegrable systems, as $E\to\overline{E}$, from above or below, a crossover between $S_A\propto N^0$ and $S_A \propto N^1$ takes place, indicating that area-law eigenstates exist only at the edges (low- and high-energies) of the spectrum.}

\change{The mean energy can be used as a reference to compute the maximum of the $S_A$'s. In fact, this maximum is easily obtained as an average $\overline{S}_A^{(\Lambda)}$ over a fixed number $(\Lambda)$ of midspectrum eigenstates. About this maximum we can write the entanglement entropy as
\begin{equation}
\label{eq:parabola_fit}
S_A = \overline{S}_A^{(\Lambda)} - C^2 (\Delta E)^2,   
\end{equation}
where $\Delta E = E - \overline{E}$ is the energy difference from the mid-energy for a given energy, and $C$ in general depends on the system Hamiltonian parameters and system size. Notice that for us $\overline{E} = {\rm tr}(H_{\rm m})/2^N = 0$ and $E = \langle H_{\rm M}\rangle$, where the expectation value is with respect to a given eigenstate.}

\change{We make use of Eq.~(\ref{eq:parabola_fit}) to investigate the nature of the parameter $C$ and as a consequence the validity of Eq.~(\ref{eq:parabola_fit}). This is done by analyzing the data in the form $\overline{S}_A^{(\Lambda)} - S_A$ vs $\Delta E$, which we show in Fig.~\ref{fig:fig_fitted_parabola}a for the four system sizes considered in the manuscript. The data collapse indicates the size-independence of the parameter $C$, which we confirm from its numerical value extracted from a linear fit to the data. Interestingly, apart from the $N=8$ system, the value of $C^2$ stabilizes to a small positive constant, and the value of the intercept obtained by the numerical fit tends to zero rapidly with increasing system size. Using the obtained value of $C$ we show the shape of the $S_A$ vs $E$ curve for the system with $N=14$ in Fig.~\ref{fig:fig_fitted_parabola}b, where a good agreement is observed.} 

\change{Finally, we investigate the scaling behavior of the mean and variance of $S_A$ for a fixed number of midspectrum eigenstates. For a chaotic system it is expected that these two quantities will converge to the prediction of the appropriate random matrix ensemble. In paritcular, the variance must decay to zero as $\sim d^{-1/2}$ where $d=2^N$ is the Hilbert space dimension. In our averages we took $\Lambda = 40$ for all system sizes. The inset of Fig.~\ref{fig:fig_fitted_parabola}a shows the results for $\overline{S}_A^{(\rm Page)} - \overline{S}_A^{(\Lambda)}$ (blue) and variance (orange), and the green solid line shows the scaling $\sim d^{-1/2}$. Clearly the mean entropy of the mispectrum eigenstates tends towards the Page value with increasing system size and the variance is going to zero with the correct scaling.}


\bibliography{references}

\begin{thebibliography}{66}%
\makeatletter
\providecommand \@ifxundefined [1]{%
 \@ifx{#1\undefined}
}%
\providecommand \@ifnum [1]{%
 \ifnum #1\expandafter \@firstoftwo
 \else \expandafter \@secondoftwo
 \fi
}%
\providecommand \@ifx [1]{%
 \ifx #1\expandafter \@firstoftwo
 \else \expandafter \@secondoftwo
 \fi
}%
\providecommand \natexlab [1]{#1}%
\providecommand \enquote  [1]{``#1''}%
\providecommand \bibnamefont  [1]{#1}%
\providecommand \bibfnamefont [1]{#1}%
\providecommand \citenamefont [1]{#1}%
\providecommand \href@noop [0]{\@secondoftwo}%
\providecommand \href [0]{\begingroup \@sanitize@url \@href}%
\providecommand \@href[1]{\@@startlink{#1}\@@href}%
\providecommand \@@href[1]{\endgroup#1\@@endlink}%
\providecommand \@sanitize@url [0]{\catcode `\\12\catcode `\$12\catcode `\&12\catcode `\#12\catcode `\^12\catcode `\_12\catcode `\%12\relax}%
\providecommand \@@startlink[1]{}%
\providecommand \@@endlink[0]{}%
\providecommand \url  [0]{\begingroup\@sanitize@url \@url }%
\providecommand \@url [1]{\endgroup\@href {#1}{\urlprefix }}%
\providecommand \urlprefix  [0]{URL }%
\providecommand \Eprint [0]{\href }%
\providecommand \doibase [0]{https://doi.org/}%
\providecommand \selectlanguage [0]{\@gobble}%
\providecommand \bibinfo  [0]{\@secondoftwo}%
\providecommand \bibfield  [0]{\@secondoftwo}%
\providecommand \translation [1]{[#1]}%
\providecommand \BibitemOpen [0]{}%
\providecommand \bibitemStop [0]{}%
\providecommand \bibitemNoStop [0]{.\EOS\space}%
\providecommand \EOS [0]{\spacefactor3000\relax}%
\providecommand \BibitemShut  [1]{\csname bibitem#1\endcsname}%
\let\auto@bib@innerbib\@empty
\bibitem [{\citenamefont {Rajak}\ \emph {et~al.}(2023)\citenamefont {Rajak}, \citenamefont {Suzuki}, \citenamefont {Dutta},\ and\ \citenamefont {Chakrabarti}}]{Rajak2023}%
  \BibitemOpen
  \bibfield  {author} {\bibinfo {author} {\bibfnamefont {A.}~\bibnamefont {Rajak}}, \bibinfo {author} {\bibfnamefont {S.}~\bibnamefont {Suzuki}}, \bibinfo {author} {\bibfnamefont {A.}~\bibnamefont {Dutta}},\ and\ \bibinfo {author} {\bibfnamefont {B.~K.}\ \bibnamefont {Chakrabarti}},\ }\bibfield  {title} {\bibinfo {title} {Quantum annealing: An overview},\ }\href {https://royalsocietypublishing.org/doi/10.1098/rsta.2021.0417} {\bibfield  {journal} {\bibinfo  {journal} {Philosophical Transactions of the Royal Society A}\ }\textbf {\bibinfo {volume} {381}},\ \bibinfo {pages} {20210417} (\bibinfo {year} {2023})}\BibitemShut {NoStop}%
\bibitem [{\citenamefont {Albash}\ and\ \citenamefont {Lidar}(2018)}]{Albash2018}%
  \BibitemOpen
  \bibfield  {author} {\bibinfo {author} {\bibfnamefont {T.}~\bibnamefont {Albash}}\ and\ \bibinfo {author} {\bibfnamefont {D.~A.}\ \bibnamefont {Lidar}},\ }\bibfield  {title} {\bibinfo {title} {Adiabatic quantum computation},\ }\href {https://doi.org/10.1103/RevModPhys.90.015002} {\bibfield  {journal} {\bibinfo  {journal} {Rev. Mod. Phys.}\ }\textbf {\bibinfo {volume} {90}},\ \bibinfo {pages} {015002} (\bibinfo {year} {2018})}\BibitemShut {NoStop}%
\bibitem [{\citenamefont {Vitanov}\ \emph {et~al.}(2017)\citenamefont {Vitanov}, \citenamefont {Rangelov}, \citenamefont {Shore},\ and\ \citenamefont {Bergmann}}]{vitanov2017}%
  \BibitemOpen
  \bibfield  {author} {\bibinfo {author} {\bibfnamefont {N.~V.}\ \bibnamefont {Vitanov}}, \bibinfo {author} {\bibfnamefont {A.~A.}\ \bibnamefont {Rangelov}}, \bibinfo {author} {\bibfnamefont {B.~W.}\ \bibnamefont {Shore}},\ and\ \bibinfo {author} {\bibfnamefont {K.}~\bibnamefont {Bergmann}},\ }\bibfield  {title} {\bibinfo {title} {Stimulated raman adiabatic passage in physics, chemistry, and beyond},\ }\href@noop {} {\bibfield  {journal} {\bibinfo  {journal} {Reviews of Modern Physics}\ }\textbf {\bibinfo {volume} {89}},\ \bibinfo {pages} {015006} (\bibinfo {year} {2017})}\BibitemShut {NoStop}%
\bibitem [{\citenamefont {Farhi}\ \emph {et~al.}(2014)\citenamefont {Farhi}, \citenamefont {Goldstone},\ and\ \citenamefont {Gutmann}}]{Farhi2014}%
  \BibitemOpen
  \bibfield  {author} {\bibinfo {author} {\bibfnamefont {E.}~\bibnamefont {Farhi}}, \bibinfo {author} {\bibfnamefont {J.}~\bibnamefont {Goldstone}},\ and\ \bibinfo {author} {\bibfnamefont {S.}~\bibnamefont {Gutmann}},\ }\bibfield  {title} {\bibinfo {title} {A quantum approximate optimization algorithm},\ }\href {https://doi.org/10.48550/arXiv.1411.4028} {\bibfield  {journal} {\bibinfo  {journal} {arXiv preprint arXiv:1411.4028}\ } (\bibinfo {year} {2014})}\BibitemShut {NoStop}%
\bibitem [{\citenamefont {Blekos}\ \emph {et~al.}(2024)\citenamefont {Blekos}, \citenamefont {Brand}, \citenamefont {Ceschini}, \citenamefont {Chou}, \citenamefont {Li}, \citenamefont {Pandya},\ and\ \citenamefont {Summer}}]{Blekos2024}%
  \BibitemOpen
  \bibfield  {author} {\bibinfo {author} {\bibfnamefont {K.}~\bibnamefont {Blekos}}, \bibinfo {author} {\bibfnamefont {D.}~\bibnamefont {Brand}}, \bibinfo {author} {\bibfnamefont {A.}~\bibnamefont {Ceschini}}, \bibinfo {author} {\bibfnamefont {C.-H.}\ \bibnamefont {Chou}}, \bibinfo {author} {\bibfnamefont {R.-H.}\ \bibnamefont {Li}}, \bibinfo {author} {\bibfnamefont {K.}~\bibnamefont {Pandya}},\ and\ \bibinfo {author} {\bibfnamefont {A.}~\bibnamefont {Summer}},\ }\bibfield  {title} {\bibinfo {title} {A review on quantum approximate optimization algorithm and its variants},\ }\href {https://www.sciencedirect.com/science/article/pii/S0370157324001078} {\bibfield  {journal} {\bibinfo  {journal} {Physics Reports}\ }\textbf {\bibinfo {volume} {1068}},\ \bibinfo {pages} {1} (\bibinfo {year} {2024})}\BibitemShut {NoStop}%
\bibitem [{\citenamefont {Rigol}\ \emph {et~al.}(2008)\citenamefont {Rigol}, \citenamefont {Dunjko},\ and\ \citenamefont {Olshanii}}]{Rigol2008}%
  \BibitemOpen
  \bibfield  {author} {\bibinfo {author} {\bibfnamefont {M.}~\bibnamefont {Rigol}}, \bibinfo {author} {\bibfnamefont {V.}~\bibnamefont {Dunjko}},\ and\ \bibinfo {author} {\bibfnamefont {M.}~\bibnamefont {Olshanii}},\ }\bibfield  {title} {\bibinfo {title} {Thermalization and its mechanism for generic isolated quantum systems},\ }\href {https://doi.org/10.1038/nature06838} {\bibfield  {journal} {\bibinfo  {journal} {Nature}\ }\textbf {\bibinfo {volume} {452}},\ \bibinfo {pages} {854} (\bibinfo {year} {2008})}\BibitemShut {NoStop}%
\bibitem [{\citenamefont {Polkovnikov}\ \emph {et~al.}(2011)\citenamefont {Polkovnikov}, \citenamefont {Sengupta}, \citenamefont {Silva},\ and\ \citenamefont {Vengalattore}}]{Polkovnikov2011}%
  \BibitemOpen
  \bibfield  {author} {\bibinfo {author} {\bibfnamefont {A.}~\bibnamefont {Polkovnikov}}, \bibinfo {author} {\bibfnamefont {K.}~\bibnamefont {Sengupta}}, \bibinfo {author} {\bibfnamefont {A.}~\bibnamefont {Silva}},\ and\ \bibinfo {author} {\bibfnamefont {M.}~\bibnamefont {Vengalattore}},\ }\bibfield  {title} {\bibinfo {title} {Colloquium: Nonequilibrium dynamics of closed interacting quantum systems},\ }\href {https://doi.org/10.1103/RevModPhys.83.863} {\bibfield  {journal} {\bibinfo  {journal} {Rev. Mod. Phys.}\ }\textbf {\bibinfo {volume} {83}},\ \bibinfo {pages} {863} (\bibinfo {year} {2011})}\BibitemShut {NoStop}%
\bibitem [{\citenamefont {Santos}\ and\ \citenamefont {Rigol}(2010)}]{Santos2010}%
  \BibitemOpen
  \bibfield  {author} {\bibinfo {author} {\bibfnamefont {L.~F.}\ \bibnamefont {Santos}}\ and\ \bibinfo {author} {\bibfnamefont {M.}~\bibnamefont {Rigol}},\ }\bibfield  {title} {\bibinfo {title} {Onset of quantum chaos in one-dimensional bosonic and fermionic systems and its relation to thermalization},\ }\href {https://doi.org/10.1103/PhysRevE.81.036206} {\bibfield  {journal} {\bibinfo  {journal} {Phys. Rev. E}\ }\textbf {\bibinfo {volume} {81}},\ \bibinfo {pages} {036206} (\bibinfo {year} {2010})}\BibitemShut {NoStop}%
\bibitem [{\citenamefont {D'Alessio}\ \emph {et~al.}(2016)\citenamefont {D'Alessio}, \citenamefont {Kafri}, \citenamefont {Polkovnikov},\ and\ \citenamefont {Rigol}}]{Dalessio2016}%
  \BibitemOpen
  \bibfield  {author} {\bibinfo {author} {\bibfnamefont {L.}~\bibnamefont {D'Alessio}}, \bibinfo {author} {\bibfnamefont {Y.}~\bibnamefont {Kafri}}, \bibinfo {author} {\bibfnamefont {A.}~\bibnamefont {Polkovnikov}},\ and\ \bibinfo {author} {\bibfnamefont {M.}~\bibnamefont {Rigol}},\ }\bibfield  {title} {\bibinfo {title} {{From quantum chaos and eigenstate thermalization to statistical mechanics and thermodynamics}},\ }\href {https://doi.org/10.1080/00018732.2016.1198134} {\bibfield  {journal} {\bibinfo  {journal} {Advances in Physics}\ }\textbf {\bibinfo {volume} {65}},\ \bibinfo {pages} {239} (\bibinfo {year} {2016})}\BibitemShut {NoStop}%
\bibitem [{\citenamefont {Mu\~noz Arias}(2022)}]{Munoz2022}%
  \BibitemOpen
  \bibfield  {author} {\bibinfo {author} {\bibfnamefont {M.~H.}\ \bibnamefont {Mu\~noz Arias}},\ }\bibfield  {title} {\bibinfo {title} {Statistical complexity and the road to equilibrium in many-body chaotic quantum systems},\ }\href {https://doi.org/10.1103/PhysRevE.106.044103} {\bibfield  {journal} {\bibinfo  {journal} {Phys. Rev. E}\ }\textbf {\bibinfo {volume} {106}},\ \bibinfo {pages} {044103} (\bibinfo {year} {2022})}\BibitemShut {NoStop}%
\bibitem [{\citenamefont {Gubin}\ and\ \citenamefont {F~Santos}(2012)}]{Gubin2012}%
  \BibitemOpen
  \bibfield  {author} {\bibinfo {author} {\bibfnamefont {A.}~\bibnamefont {Gubin}}\ and\ \bibinfo {author} {\bibfnamefont {L.}~\bibnamefont {F~Santos}},\ }\bibfield  {title} {\bibinfo {title} {Quantum chaos: An introduction via chains of interacting spins 1/2},\ }\href {https://pubs.aip.org/aapt/ajp/article-abstract/80/3/246/1056816/Quantum-chaos-An-introduction-via-chains-of?redirectedFrom=fulltext} {\bibfield  {journal} {\bibinfo  {journal} {American Journal of Physics}\ }\textbf {\bibinfo {volume} {80}},\ \bibinfo {pages} {246} (\bibinfo {year} {2012})}\BibitemShut {NoStop}%
\bibitem [{\citenamefont {\L{}yd\ifmmode~\dot{z}\else \.{z}\fi{}ba}\ \emph {et~al.}(2021)\citenamefont {\L{}yd\ifmmode~\dot{z}\else \.{z}\fi{}ba}, \citenamefont {Rigol},\ and\ \citenamefont {Vidmar}}]{Lydzba2021}%
  \BibitemOpen
  \bibfield  {author} {\bibinfo {author} {\bibfnamefont {P.}~\bibnamefont {\L{}yd\ifmmode~\dot{z}\else \.{z}\fi{}ba}}, \bibinfo {author} {\bibfnamefont {M.}~\bibnamefont {Rigol}},\ and\ \bibinfo {author} {\bibfnamefont {L.}~\bibnamefont {Vidmar}},\ }\bibfield  {title} {\bibinfo {title} {Entanglement in many-body eigenstates of quantum-chaotic quadratic hamiltonians},\ }\href {https://doi.org/10.1103/PhysRevB.103.104206} {\bibfield  {journal} {\bibinfo  {journal} {Phys. Rev. B}\ }\textbf {\bibinfo {volume} {103}},\ \bibinfo {pages} {104206} (\bibinfo {year} {2021})}\BibitemShut {NoStop}%
\bibitem [{\citenamefont {Kliczkowski}\ \emph {et~al.}(2023)\citenamefont {Kliczkowski}, \citenamefont {\'Swi{\k e}tek}, \citenamefont {Vidmar},\ and\ \citenamefont {Rigol}}]{Kliczkowski2023}%
  \BibitemOpen
  \bibfield  {author} {\bibinfo {author} {\bibfnamefont {M.}~\bibnamefont {Kliczkowski}}, \bibinfo {author} {\bibfnamefont {R.}~\bibnamefont {\'Swi{\k e}tek}}, \bibinfo {author} {\bibfnamefont {L.}~\bibnamefont {Vidmar}},\ and\ \bibinfo {author} {\bibfnamefont {M.}~\bibnamefont {Rigol}},\ }\bibfield  {title} {\bibinfo {title} {Average entanglement entropy of midspectrum eigenstates of quantum-chaotic interacting hamiltonians},\ }\href {https://doi.org/10.1103/PhysRevE.107.064119} {\bibfield  {journal} {\bibinfo  {journal} {Phys. Rev. E}\ }\textbf {\bibinfo {volume} {107}},\ \bibinfo {pages} {064119} (\bibinfo {year} {2023})}\BibitemShut {NoStop}%
\bibitem [{\citenamefont {Wang}\ \emph {et~al.}(2004)\citenamefont {Wang}, \citenamefont {Ghose}, \citenamefont {Sanders},\ and\ \citenamefont {Hu}}]{wang2004}%
  \BibitemOpen
  \bibfield  {author} {\bibinfo {author} {\bibfnamefont {X.}~\bibnamefont {Wang}}, \bibinfo {author} {\bibfnamefont {S.}~\bibnamefont {Ghose}}, \bibinfo {author} {\bibfnamefont {B.~C.}\ \bibnamefont {Sanders}},\ and\ \bibinfo {author} {\bibfnamefont {B.}~\bibnamefont {Hu}},\ }\bibfield  {title} {\bibinfo {title} {Entanglement as a signature of quantum chaos},\ }\href@noop {} {\bibfield  {journal} {\bibinfo  {journal} {Physical Review E—Statistical, Nonlinear, and Soft Matter Physics}\ }\textbf {\bibinfo {volume} {70}},\ \bibinfo {pages} {016217} (\bibinfo {year} {2004})}\BibitemShut {NoStop}%
\bibitem [{\citenamefont {Chang}\ \emph {et~al.}(2019)\citenamefont {Chang}, \citenamefont {Chen}, \citenamefont {Gopalakrishnan},\ and\ \citenamefont {Pixley}}]{Chang2019}%
  \BibitemOpen
  \bibfield  {author} {\bibinfo {author} {\bibfnamefont {P.-Y.}\ \bibnamefont {Chang}}, \bibinfo {author} {\bibfnamefont {X.}~\bibnamefont {Chen}}, \bibinfo {author} {\bibfnamefont {S.}~\bibnamefont {Gopalakrishnan}},\ and\ \bibinfo {author} {\bibfnamefont {J.~H.}\ \bibnamefont {Pixley}},\ }\bibfield  {title} {\bibinfo {title} {Evolution of entanglement spectra under generic quantum dynamics},\ }\href {https://doi.org/10.1103/PhysRevLett.123.190602} {\bibfield  {journal} {\bibinfo  {journal} {Phys. Rev. Lett.}\ }\textbf {\bibinfo {volume} {123}},\ \bibinfo {pages} {190602} (\bibinfo {year} {2019})}\BibitemShut {NoStop}%
\bibitem [{\citenamefont {D'Alessio}\ and\ \citenamefont {Rigol}(2014)}]{Dalessio2014}%
  \BibitemOpen
  \bibfield  {author} {\bibinfo {author} {\bibfnamefont {L.}~\bibnamefont {D'Alessio}}\ and\ \bibinfo {author} {\bibfnamefont {M.}~\bibnamefont {Rigol}},\ }\bibfield  {title} {\bibinfo {title} {Long-time behavior of isolated periodically driven interacting lattice systems},\ }\href {https://doi.org/10.1103/PhysRevX.4.041048} {\bibfield  {journal} {\bibinfo  {journal} {Phys. Rev. X}\ }\textbf {\bibinfo {volume} {4}},\ \bibinfo {pages} {041048} (\bibinfo {year} {2014})}\BibitemShut {NoStop}%
\bibitem [{\citenamefont {Haake}\ \emph {et~al.}(1987)\citenamefont {Haake}, \citenamefont {Ku{\'s}},\ and\ \citenamefont {Scharf}}]{haake1987}%
  \BibitemOpen
  \bibfield  {author} {\bibinfo {author} {\bibfnamefont {F.}~\bibnamefont {Haake}}, \bibinfo {author} {\bibfnamefont {M.}~\bibnamefont {Ku{\'s}}},\ and\ \bibinfo {author} {\bibfnamefont {R.}~\bibnamefont {Scharf}},\ }\bibfield  {title} {\bibinfo {title} {Classical and quantum chaos for a kicked top},\ }\href@noop {} {\bibfield  {journal} {\bibinfo  {journal} {Zeitschrift f{\"u}r Physik B Condensed Matter}\ }\textbf {\bibinfo {volume} {65}},\ \bibinfo {pages} {381} (\bibinfo {year} {1987})}\BibitemShut {NoStop}%
\bibitem [{\citenamefont {Page}(1993)}]{Page1993}%
  \BibitemOpen
  \bibfield  {author} {\bibinfo {author} {\bibfnamefont {D.~N.}\ \bibnamefont {Page}},\ }\bibfield  {title} {\bibinfo {title} {Average entropy of a subsystem},\ }\href {https://doi.org/10.1103/PhysRevLett.71.1291} {\bibfield  {journal} {\bibinfo  {journal} {Phys. Rev. Lett.}\ }\textbf {\bibinfo {volume} {71}},\ \bibinfo {pages} {1291} (\bibinfo {year} {1993})}\BibitemShut {NoStop}%
\bibitem [{\citenamefont {Lucas}(2014)}]{Lucas2014}%
  \BibitemOpen
  \bibfield  {author} {\bibinfo {author} {\bibfnamefont {A.}~\bibnamefont {Lucas}},\ }\bibfield  {title} {\bibinfo {title} {Ising formulations of many np problems},\ }\bibfield  {journal} {\bibinfo  {journal} {Frontiers in Physics}\ }\textbf {\bibinfo {volume} {2}},\ \href {https://doi.org/10.3389/fphy.2014.00005} {10.3389/fphy.2014.00005} (\bibinfo {year} {2014})\BibitemShut {NoStop}%
\bibitem [{\citenamefont {Ebadi}\ \emph {et~al.}(2022)\citenamefont {Ebadi}, \citenamefont {Keesling}, \citenamefont {Cain}, \citenamefont {Wang}, \citenamefont {Levine}, \citenamefont {Bluvstein}, \citenamefont {Semeghini}, \citenamefont {Omran}, \citenamefont {Liu}, \citenamefont {Samajdar}, \citenamefont {Luo}, \citenamefont {Nash}, \citenamefont {Gao}, \citenamefont {Barak}, \citenamefont {Farhi}, \citenamefont {Sachdev}, \citenamefont {Gemelke}, \citenamefont {Zhou}, \citenamefont {Choi}, \citenamefont {Pichler}, \citenamefont {Wang}, \citenamefont {Greiner}, \citenamefont {Vuletić},\ and\ \citenamefont {Lukin}}]{Ebadi2022}%
  \BibitemOpen
  \bibfield  {author} {\bibinfo {author} {\bibfnamefont {S.}~\bibnamefont {Ebadi}}, \bibinfo {author} {\bibfnamefont {A.}~\bibnamefont {Keesling}}, \bibinfo {author} {\bibfnamefont {M.}~\bibnamefont {Cain}}, \bibinfo {author} {\bibfnamefont {T.~T.}\ \bibnamefont {Wang}}, \bibinfo {author} {\bibfnamefont {H.}~\bibnamefont {Levine}}, \bibinfo {author} {\bibfnamefont {D.}~\bibnamefont {Bluvstein}}, \bibinfo {author} {\bibfnamefont {G.}~\bibnamefont {Semeghini}}, \bibinfo {author} {\bibfnamefont {A.}~\bibnamefont {Omran}}, \bibinfo {author} {\bibfnamefont {J.-G.}\ \bibnamefont {Liu}}, \bibinfo {author} {\bibfnamefont {R.}~\bibnamefont {Samajdar}}, \bibinfo {author} {\bibfnamefont {X.-Z.}\ \bibnamefont {Luo}}, \bibinfo {author} {\bibfnamefont {B.}~\bibnamefont {Nash}}, \bibinfo {author} {\bibfnamefont {X.}~\bibnamefont {Gao}}, \bibinfo {author} {\bibfnamefont {B.}~\bibnamefont {Barak}}, \bibinfo {author} {\bibfnamefont {E.}~\bibnamefont {Farhi}}, \bibinfo {author} {\bibfnamefont {S.}~\bibnamefont {Sachdev}},
  \bibinfo {author} {\bibfnamefont {N.}~\bibnamefont {Gemelke}}, \bibinfo {author} {\bibfnamefont {L.}~\bibnamefont {Zhou}}, \bibinfo {author} {\bibfnamefont {S.}~\bibnamefont {Choi}}, \bibinfo {author} {\bibfnamefont {H.}~\bibnamefont {Pichler}}, \bibinfo {author} {\bibfnamefont {S.-T.}\ \bibnamefont {Wang}}, \bibinfo {author} {\bibfnamefont {M.}~\bibnamefont {Greiner}}, \bibinfo {author} {\bibfnamefont {V.}~\bibnamefont {Vuletić}},\ and\ \bibinfo {author} {\bibfnamefont {M.~D.}\ \bibnamefont {Lukin}},\ }\bibfield  {title} {\bibinfo {title} {Quantum optimization of maximum independent set using rydberg atom arrays},\ }\href {https://doi.org/10.1126/science.abo6587} {\bibfield  {journal} {\bibinfo  {journal} {Science}\ }\textbf {\bibinfo {volume} {376}},\ \bibinfo {pages} {1209} (\bibinfo {year} {2022})},\ \Eprint {https://arxiv.org/abs/https://www.science.org/doi/pdf/10.1126/science.abo6587} {https://www.science.org/doi/pdf/10.1126/science.abo6587} \BibitemShut {NoStop}%
\bibitem [{\citenamefont {Sherrington}\ and\ \citenamefont {Kirkpatrick}(1975)}]{Sherrington1975}%
  \BibitemOpen
  \bibfield  {author} {\bibinfo {author} {\bibfnamefont {D.}~\bibnamefont {Sherrington}}\ and\ \bibinfo {author} {\bibfnamefont {S.}~\bibnamefont {Kirkpatrick}},\ }\bibfield  {title} {\bibinfo {title} {Solvable model of a spin-glass},\ }\href {https://doi.org/10.1103/PhysRevLett.35.1792} {\bibfield  {journal} {\bibinfo  {journal} {Phys. Rev. Lett.}\ }\textbf {\bibinfo {volume} {35}},\ \bibinfo {pages} {1792} (\bibinfo {year} {1975})}\BibitemShut {NoStop}%
\bibitem [{\citenamefont {Charbonneau}\ \emph {et~al.}(2023)\citenamefont {Charbonneau}, \citenamefont {Marinari}, \citenamefont {Mézard}, \citenamefont {Parisi}, \citenamefont {Ricci-Tersenghi}, \citenamefont {Sicuro},\ and\ \citenamefont {Zamponi}}]{Charbonneau2023}%
  \BibitemOpen
  \bibfield  {author} {\bibinfo {author} {\bibfnamefont {P.}~\bibnamefont {Charbonneau}}, \bibinfo {author} {\bibfnamefont {E.}~\bibnamefont {Marinari}}, \bibinfo {author} {\bibfnamefont {M.}~\bibnamefont {Mézard}}, \bibinfo {author} {\bibfnamefont {G.}~\bibnamefont {Parisi}}, \bibinfo {author} {\bibfnamefont {F.}~\bibnamefont {Ricci-Tersenghi}}, \bibinfo {author} {\bibfnamefont {G.}~\bibnamefont {Sicuro}},\ and\ \bibinfo {author} {\bibfnamefont {F.}~\bibnamefont {Zamponi}},\ }\href {https://doi.org/10.1142/13341} {\emph {\bibinfo {title} {Spin Glass Theory and Far Beyond}}}\ (\bibinfo  {publisher} {WORLD SCIENTIFIC},\ \bibinfo {year} {2023})\ \Eprint {https://arxiv.org/abs/https://www.worldscientific.com/doi/pdf/10.1142/13341} {https://www.worldscientific.com/doi/pdf/10.1142/13341} \BibitemShut {NoStop}%
\bibitem [{\citenamefont {Karthik}\ \emph {et~al.}(2007)\citenamefont {Karthik}, \citenamefont {Sharma},\ and\ \citenamefont {Lakshminarayan}}]{Karthik2007}%
  \BibitemOpen
  \bibfield  {author} {\bibinfo {author} {\bibfnamefont {J.}~\bibnamefont {Karthik}}, \bibinfo {author} {\bibfnamefont {A.}~\bibnamefont {Sharma}},\ and\ \bibinfo {author} {\bibfnamefont {A.}~\bibnamefont {Lakshminarayan}},\ }\bibfield  {title} {\bibinfo {title} {Entanglement, avoided crossings, and quantum chaos in an ising model with a tilted magnetic field},\ }\href {https://doi.org/10.1103/PhysRevA.75.022304} {\bibfield  {journal} {\bibinfo  {journal} {Phys. Rev. A}\ }\textbf {\bibinfo {volume} {75}},\ \bibinfo {pages} {022304} (\bibinfo {year} {2007})}\BibitemShut {NoStop}%
\bibitem [{\citenamefont {Kim}\ and\ \citenamefont {Huse}(2013)}]{Kim2013}%
  \BibitemOpen
  \bibfield  {author} {\bibinfo {author} {\bibfnamefont {H.}~\bibnamefont {Kim}}\ and\ \bibinfo {author} {\bibfnamefont {D.~A.}\ \bibnamefont {Huse}},\ }\bibfield  {title} {\bibinfo {title} {Ballistic spreading of entanglement in a diffusive nonintegrable system},\ }\href {https://doi.org/10.1103/PhysRevLett.111.127205} {\bibfield  {journal} {\bibinfo  {journal} {Phys. Rev. Lett.}\ }\textbf {\bibinfo {volume} {111}},\ \bibinfo {pages} {127205} (\bibinfo {year} {2013})}\BibitemShut {NoStop}%
\bibitem [{\citenamefont {Kim}\ \emph {et~al.}(2015)\citenamefont {Kim}, \citenamefont {Ba\~nuls}, \citenamefont {Cirac}, \citenamefont {Hastings},\ and\ \citenamefont {Huse}}]{Kim2015}%
  \BibitemOpen
  \bibfield  {author} {\bibinfo {author} {\bibfnamefont {H.}~\bibnamefont {Kim}}, \bibinfo {author} {\bibfnamefont {M.~C.}\ \bibnamefont {Ba\~nuls}}, \bibinfo {author} {\bibfnamefont {J.~I.}\ \bibnamefont {Cirac}}, \bibinfo {author} {\bibfnamefont {M.~B.}\ \bibnamefont {Hastings}},\ and\ \bibinfo {author} {\bibfnamefont {D.~A.}\ \bibnamefont {Huse}},\ }\bibfield  {title} {\bibinfo {title} {Slowest local operators in quantum spin chains},\ }\href {https://doi.org/10.1103/PhysRevE.92.012128} {\bibfield  {journal} {\bibinfo  {journal} {Phys. Rev. E}\ }\textbf {\bibinfo {volume} {92}},\ \bibinfo {pages} {012128} (\bibinfo {year} {2015})}\BibitemShut {NoStop}%
\bibitem [{\citenamefont {Grabarits}\ \emph {et~al.}(2024)\citenamefont {Grabarits}, \citenamefont {Swain}, \citenamefont {Heydari}, \citenamefont {Chandarana}, \citenamefont {G{\'o}mez-Ruiz},\ and\ \citenamefont {del Campo}}]{grabarits2024}%
  \BibitemOpen
  \bibfield  {author} {\bibinfo {author} {\bibfnamefont {A.}~\bibnamefont {Grabarits}}, \bibinfo {author} {\bibfnamefont {K.~R.}\ \bibnamefont {Swain}}, \bibinfo {author} {\bibfnamefont {M.~S.}\ \bibnamefont {Heydari}}, \bibinfo {author} {\bibfnamefont {P.}~\bibnamefont {Chandarana}}, \bibinfo {author} {\bibfnamefont {F.~J.}\ \bibnamefont {G{\'o}mez-Ruiz}},\ and\ \bibinfo {author} {\bibfnamefont {A.}~\bibnamefont {del Campo}},\ }\bibfield  {title} {\bibinfo {title} {Quantum chaos in random ising networks},\ }\href@noop {} {\bibfield  {journal} {\bibinfo  {journal} {arXiv preprint arXiv:2405.14376}\ } (\bibinfo {year} {2024})}\BibitemShut {NoStop}%
\bibitem [{\citenamefont {Atas}\ \emph {et~al.}(2013)\citenamefont {Atas}, \citenamefont {Bogomolny}, \citenamefont {Giraud},\ and\ \citenamefont {Roux}}]{Atas2013}%
  \BibitemOpen
  \bibfield  {author} {\bibinfo {author} {\bibfnamefont {Y.~Y.}\ \bibnamefont {Atas}}, \bibinfo {author} {\bibfnamefont {E.}~\bibnamefont {Bogomolny}}, \bibinfo {author} {\bibfnamefont {O.}~\bibnamefont {Giraud}},\ and\ \bibinfo {author} {\bibfnamefont {G.}~\bibnamefont {Roux}},\ }\bibfield  {title} {\bibinfo {title} {Distribution of the ratio of consecutive level spacings in random matrix ensembles},\ }\href {https://doi.org/10.1103/PhysRevLett.110.084101} {\bibfield  {journal} {\bibinfo  {journal} {Phys. Rev. Lett.}\ }\textbf {\bibinfo {volume} {110}},\ \bibinfo {pages} {084101} (\bibinfo {year} {2013})}\BibitemShut {NoStop}%
\bibitem [{\citenamefont {Berry}(1984)}]{Berry1984}%
  \BibitemOpen
  \bibfield  {author} {\bibinfo {author} {\bibfnamefont {M.~V.}\ \bibnamefont {Berry}},\ }\bibfield  {title} {\bibinfo {title} {Quantal phase factors accompanying adiabatic changes},\ }\href {https://doi.org/10.1098/rspa.1984.0023} {\bibfield  {journal} {\bibinfo  {journal} {Proc. R. Soc. Lond. A}\ }\textbf {\bibinfo {volume} {392}},\ \bibinfo {pages} {45–57} (\bibinfo {year} {1984})}\BibitemShut {NoStop}%
\bibitem [{\citenamefont {Born}\ and\ \citenamefont {Fock}(1928)}]{Born1928}%
  \BibitemOpen
  \bibfield  {author} {\bibinfo {author} {\bibfnamefont {M.}~\bibnamefont {Born}}\ and\ \bibinfo {author} {\bibfnamefont {V.}~\bibnamefont {Fock}},\ }\bibfield  {title} {\bibinfo {title} {Beweis des adiabatensatzes},\ }\href {https://doi.org/10.1007/bf01343193} {\bibfield  {journal} {\bibinfo  {journal} {Zeitschrift f\"ur Physik}\ }\textbf {\bibinfo {volume} {51}},\ \bibinfo {pages} {165–180} (\bibinfo {year} {1928})}\BibitemShut {NoStop}%
\bibitem [{\citenamefont {Kato}(1950)}]{Kato1950}%
  \BibitemOpen
  \bibfield  {author} {\bibinfo {author} {\bibfnamefont {T.}~\bibnamefont {Kato}},\ }\bibfield  {title} {\bibinfo {title} {On the adiabatic theorem of quantum mechanics},\ }\href {https://doi.org/10.1143/JPSJ.5.435} {\bibfield  {journal} {\bibinfo  {journal} {Journal of the Physical Society of Japan}\ }\textbf {\bibinfo {volume} {5}},\ \bibinfo {pages} {435} (\bibinfo {year} {1950})}\BibitemShut {NoStop}%
\bibitem [{\citenamefont {Yarloo}\ \emph {et~al.}(2024)\citenamefont {Yarloo}, \citenamefont {Zhang},\ and\ \citenamefont {Nielsen}}]{Yarloo2024}%
  \BibitemOpen
  \bibfield  {author} {\bibinfo {author} {\bibfnamefont {H.}~\bibnamefont {Yarloo}}, \bibinfo {author} {\bibfnamefont {H.-C.}\ \bibnamefont {Zhang}},\ and\ \bibinfo {author} {\bibfnamefont {A.~E.~B.}\ \bibnamefont {Nielsen}},\ }\bibfield  {title} {\bibinfo {title} {Adiabatic time evolution of highly excited states},\ }\href {https://doi.org/10.1103/PRXQuantum.5.020365} {\bibfield  {journal} {\bibinfo  {journal} {PRX Quantum}\ }\textbf {\bibinfo {volume} {5}},\ \bibinfo {pages} {020365} (\bibinfo {year} {2024})}\BibitemShut {NoStop}%
\bibitem [{\citenamefont {Puebla}\ and\ \citenamefont {G{\'o}mez-Ruiz}(2024)}]{Puebla2024}%
  \BibitemOpen
  \bibfield  {author} {\bibinfo {author} {\bibfnamefont {R.}~\bibnamefont {Puebla}}\ and\ \bibinfo {author} {\bibfnamefont {F.~J.}\ \bibnamefont {G{\'o}mez-Ruiz}},\ }\bibfield  {title} {\bibinfo {title} {Quantum information scrambling in adiabatically driven critical systems},\ }\href@noop {} {\bibfield  {journal} {\bibinfo  {journal} {Entropy}\ }\textbf {\bibinfo {volume} {26}},\ \bibinfo {pages} {951} (\bibinfo {year} {2024})}\BibitemShut {NoStop}%
\bibitem [{\citenamefont {Jansen}\ \emph {et~al.}(2007)\citenamefont {Jansen}, \citenamefont {Ruskai},\ and\ \citenamefont {Seiler}}]{Jansen2007}%
  \BibitemOpen
  \bibfield  {author} {\bibinfo {author} {\bibfnamefont {S.}~\bibnamefont {Jansen}}, \bibinfo {author} {\bibfnamefont {M.-B.}\ \bibnamefont {Ruskai}},\ and\ \bibinfo {author} {\bibfnamefont {R.}~\bibnamefont {Seiler}},\ }\bibfield  {title} {\bibinfo {title} {Bounds for the adiabatic approximation with applications to quantum computation},\ }\href {https://pubs.aip.org/aip/jmp/article-abstract/48/10/102111/379272/Bounds-for-the-adiabatic-approximation-with?redirectedFrom=fulltext} {\bibfield  {journal} {\bibinfo  {journal} {Journal of Mathematical Physics}\ }\textbf {\bibinfo {volume} {48}} (\bibinfo {year} {2007})}\BibitemShut {NoStop}%
\bibitem [{\citenamefont {Schiffer}\ \emph {et~al.}(2024)\citenamefont {Schiffer}, \citenamefont {Rubio}, \citenamefont {Trivedi},\ and\ \citenamefont {Cirac}}]{Schiffer2024}%
  \BibitemOpen
  \bibfield  {author} {\bibinfo {author} {\bibfnamefont {B.~F.}\ \bibnamefont {Schiffer}}, \bibinfo {author} {\bibfnamefont {A.~F.}\ \bibnamefont {Rubio}}, \bibinfo {author} {\bibfnamefont {R.}~\bibnamefont {Trivedi}},\ and\ \bibinfo {author} {\bibfnamefont {J.~I.}\ \bibnamefont {Cirac}},\ }\bibfield  {title} {\bibinfo {title} {The quantum adiabatic algorithm suppresses the proliferation of errors},\ }\href@noop {} {\bibfield  {journal} {\bibinfo  {journal} {arXiv preprint arXiv:2404.15397}\ } (\bibinfo {year} {2024})}\BibitemShut {NoStop}%
\bibitem [{\citenamefont {Yi}(2021)}]{Yi2021}%
  \BibitemOpen
  \bibfield  {author} {\bibinfo {author} {\bibfnamefont {C.}~\bibnamefont {Yi}},\ }\bibfield  {title} {\bibinfo {title} {Success of digital adiabatic simulation with large trotter step},\ }\href {https://doi.org/10.1103/PhysRevA.104.052603} {\bibfield  {journal} {\bibinfo  {journal} {Phys. Rev. A}\ }\textbf {\bibinfo {volume} {104}},\ \bibinfo {pages} {052603} (\bibinfo {year} {2021})}\BibitemShut {NoStop}%
\bibitem [{\citenamefont {Kovalsky}\ \emph {et~al.}(2023)\citenamefont {Kovalsky}, \citenamefont {Calderon-Vargas}, \citenamefont {Grace}, \citenamefont {Magann}, \citenamefont {Larsen}, \citenamefont {Baczewski},\ and\ \citenamefont {Sarovar}}]{Kovalsky2023}%
  \BibitemOpen
  \bibfield  {author} {\bibinfo {author} {\bibfnamefont {L.~K.}\ \bibnamefont {Kovalsky}}, \bibinfo {author} {\bibfnamefont {F.~A.}\ \bibnamefont {Calderon-Vargas}}, \bibinfo {author} {\bibfnamefont {M.~D.}\ \bibnamefont {Grace}}, \bibinfo {author} {\bibfnamefont {A.~B.}\ \bibnamefont {Magann}}, \bibinfo {author} {\bibfnamefont {J.~B.}\ \bibnamefont {Larsen}}, \bibinfo {author} {\bibfnamefont {A.~D.}\ \bibnamefont {Baczewski}},\ and\ \bibinfo {author} {\bibfnamefont {M.}~\bibnamefont {Sarovar}},\ }\bibfield  {title} {\bibinfo {title} {Self-healing of trotter error in digital adiabatic state preparation},\ }\href {https://doi.org/10.1103/PhysRevLett.131.060602} {\bibfield  {journal} {\bibinfo  {journal} {Phys. Rev. Lett.}\ }\textbf {\bibinfo {volume} {131}},\ \bibinfo {pages} {060602} (\bibinfo {year} {2023})}\BibitemShut {NoStop}%
\bibitem [{\citenamefont {Radcliffe}(1971)}]{Radcliffe1971}%
  \BibitemOpen
  \bibfield  {author} {\bibinfo {author} {\bibfnamefont {J.~M.}\ \bibnamefont {Radcliffe}},\ }\bibfield  {title} {\bibinfo {title} {Some properties of coherent spin states},\ }\href {https://doi.org/10.1088/0305-4470/4/3/009} {\bibfield  {journal} {\bibinfo  {journal} {Journal of Physics A: General Physics}\ }\textbf {\bibinfo {volume} {4}},\ \bibinfo {pages} {313} (\bibinfo {year} {1971})}\BibitemShut {NoStop}%
\bibitem [{Note1()}]{Note1}%
  \BibitemOpen
  \bibinfo {note} {Notice that in absence of the longitudinal field $\DOTSB \sum@ \slimits@ _i \sigma _i^z$ the two states $|\uparrow \rangle ^{\otimes N}$ and $|\downarrow \rangle ^{\otimes N}$ are degenerate with energy $E = -(N-1)$, thus the ground state of $-H_{\protect \rm P}$ corresponds to their even and odd superpositions. However, the inclusion of the longitudinal term shifts $|\uparrow \rangle ^{\otimes N}$ down in energy to $E = -2N + 1$ and $|\downarrow \rangle ^{\otimes N}$ up in energy to $E = 1$ effectively lifting the degeneracy, leaving $|\uparrow \rangle ^{\otimes N}$ as the unique ground state.}\BibitemShut {Stop}%
\bibitem [{\citenamefont {Haake}(1991)}]{haake1991}%
  \BibitemOpen
  \bibfield  {author} {\bibinfo {author} {\bibfnamefont {F.}~\bibnamefont {Haake}},\ }\href@noop {} {\emph {\bibinfo {title} {Quantum signatures of chaos}}}\ (\bibinfo  {publisher} {Springer},\ \bibinfo {year} {1991})\BibitemShut {NoStop}%
\bibitem [{\citenamefont {Mag\'an}(2016)}]{Magan2016}%
  \BibitemOpen
  \bibfield  {author} {\bibinfo {author} {\bibfnamefont {J.~M.}\ \bibnamefont {Mag\'an}},\ }\bibfield  {title} {\bibinfo {title} {Random free fermions: An analytical example of eigenstate thermalization},\ }\href {https://doi.org/10.1103/PhysRevLett.116.030401} {\bibfield  {journal} {\bibinfo  {journal} {Phys. Rev. Lett.}\ }\textbf {\bibinfo {volume} {116}},\ \bibinfo {pages} {030401} (\bibinfo {year} {2016})}\BibitemShut {NoStop}%
\bibitem [{\citenamefont {Bianchi}\ \emph {et~al.}(2021)\citenamefont {Bianchi}, \citenamefont {Hackl},\ and\ \citenamefont {Kieburg}}]{Bianchi2021}%
  \BibitemOpen
  \bibfield  {author} {\bibinfo {author} {\bibfnamefont {E.}~\bibnamefont {Bianchi}}, \bibinfo {author} {\bibfnamefont {L.}~\bibnamefont {Hackl}},\ and\ \bibinfo {author} {\bibfnamefont {M.}~\bibnamefont {Kieburg}},\ }\bibfield  {title} {\bibinfo {title} {Page curve for fermionic gaussian states},\ }\href {https://doi.org/10.1103/PhysRevB.103.L241118} {\bibfield  {journal} {\bibinfo  {journal} {Phys. Rev. B}\ }\textbf {\bibinfo {volume} {103}},\ \bibinfo {pages} {L241118} (\bibinfo {year} {2021})}\BibitemShut {NoStop}%
\bibitem [{\citenamefont {Amico}\ \emph {et~al.}(2008)\citenamefont {Amico}, \citenamefont {Fazio}, \citenamefont {Osterloh},\ and\ \citenamefont {Vedral}}]{Amico2008}%
  \BibitemOpen
  \bibfield  {author} {\bibinfo {author} {\bibfnamefont {L.}~\bibnamefont {Amico}}, \bibinfo {author} {\bibfnamefont {R.}~\bibnamefont {Fazio}}, \bibinfo {author} {\bibfnamefont {A.}~\bibnamefont {Osterloh}},\ and\ \bibinfo {author} {\bibfnamefont {V.}~\bibnamefont {Vedral}},\ }\bibfield  {title} {\bibinfo {title} {Entanglement in many-body systems},\ }\href {https://doi.org/10.1103/RevModPhys.80.517} {\bibfield  {journal} {\bibinfo  {journal} {Rev. Mod. Phys.}\ }\textbf {\bibinfo {volume} {80}},\ \bibinfo {pages} {517} (\bibinfo {year} {2008})}\BibitemShut {NoStop}%
\bibitem [{\citenamefont {Bianchi}\ \emph {et~al.}(2022)\citenamefont {Bianchi}, \citenamefont {Hackl}, \citenamefont {Kieburg}, \citenamefont {Rigol},\ and\ \citenamefont {Vidmar}}]{Bianchi2022}%
  \BibitemOpen
  \bibfield  {author} {\bibinfo {author} {\bibfnamefont {E.}~\bibnamefont {Bianchi}}, \bibinfo {author} {\bibfnamefont {L.}~\bibnamefont {Hackl}}, \bibinfo {author} {\bibfnamefont {M.}~\bibnamefont {Kieburg}}, \bibinfo {author} {\bibfnamefont {M.}~\bibnamefont {Rigol}},\ and\ \bibinfo {author} {\bibfnamefont {L.}~\bibnamefont {Vidmar}},\ }\bibfield  {title} {\bibinfo {title} {Volume-law entanglement entropy of typical pure quantum states},\ }\href {https://doi.org/10.1103/PRXQuantum.3.030201} {\bibfield  {journal} {\bibinfo  {journal} {PRX Quantum}\ }\textbf {\bibinfo {volume} {3}},\ \bibinfo {pages} {030201} (\bibinfo {year} {2022})}\BibitemShut {NoStop}%
\bibitem [{\citenamefont {Beugeling}\ \emph {et~al.}(2015)\citenamefont {Beugeling}, \citenamefont {Andreanov},\ and\ \citenamefont {Haque}}]{Beugeling2015}%
  \BibitemOpen
  \bibfield  {author} {\bibinfo {author} {\bibfnamefont {W.}~\bibnamefont {Beugeling}}, \bibinfo {author} {\bibfnamefont {A.}~\bibnamefont {Andreanov}},\ and\ \bibinfo {author} {\bibfnamefont {M.}~\bibnamefont {Haque}},\ }\bibfield  {title} {\bibinfo {title} {Global characteristics of all eigenstates of local many-body hamiltonians: participation ratio and entanglement entropy},\ }\href@noop {} {\bibfield  {journal} {\bibinfo  {journal} {Journal of Statistical Mechanics: Theory and Experiment}\ }\textbf {\bibinfo {volume} {2015}},\ \bibinfo {pages} {P02002} (\bibinfo {year} {2015})}\BibitemShut {NoStop}%
\bibitem [{\citenamefont {Miao}\ and\ \citenamefont {Barthel}(2021)}]{Miao2021}%
  \BibitemOpen
  \bibfield  {author} {\bibinfo {author} {\bibfnamefont {Q.}~\bibnamefont {Miao}}\ and\ \bibinfo {author} {\bibfnamefont {T.}~\bibnamefont {Barthel}},\ }\bibfield  {title} {\bibinfo {title} {Eigenstate entanglement: Crossover from the ground state to volume laws},\ }\href {https://doi.org/10.1103/PhysRevLett.127.040603} {\bibfield  {journal} {\bibinfo  {journal} {Phys. Rev. Lett.}\ }\textbf {\bibinfo {volume} {127}},\ \bibinfo {pages} {040603} (\bibinfo {year} {2021})}\BibitemShut {NoStop}%
\bibitem [{\citenamefont {Serbyn}\ \emph {et~al.}(2021)\citenamefont {Serbyn}, \citenamefont {Abanin},\ and\ \citenamefont {Papi{\'c}}}]{serbyn2021}%
  \BibitemOpen
  \bibfield  {author} {\bibinfo {author} {\bibfnamefont {M.}~\bibnamefont {Serbyn}}, \bibinfo {author} {\bibfnamefont {D.~A.}\ \bibnamefont {Abanin}},\ and\ \bibinfo {author} {\bibfnamefont {Z.}~\bibnamefont {Papi{\'c}}},\ }\bibfield  {title} {\bibinfo {title} {Quantum many-body scars and weak breaking of ergodicity},\ }\href@noop {} {\bibfield  {journal} {\bibinfo  {journal} {Nature Physics}\ }\textbf {\bibinfo {volume} {17}},\ \bibinfo {pages} {675} (\bibinfo {year} {2021})}\BibitemShut {NoStop}%
\bibitem [{\citenamefont {Garrison}\ and\ \citenamefont {Grover}(2018)}]{Garrison2018}%
  \BibitemOpen
  \bibfield  {author} {\bibinfo {author} {\bibfnamefont {J.~R.}\ \bibnamefont {Garrison}}\ and\ \bibinfo {author} {\bibfnamefont {T.}~\bibnamefont {Grover}},\ }\bibfield  {title} {\bibinfo {title} {Does a single eigenstate encode the full hamiltonian?},\ }\href {https://doi.org/10.1103/PhysRevX.8.021026} {\bibfield  {journal} {\bibinfo  {journal} {Phys. Rev. X}\ }\textbf {\bibinfo {volume} {8}},\ \bibinfo {pages} {021026} (\bibinfo {year} {2018})}\BibitemShut {NoStop}%
\bibitem [{\citenamefont {Haque}\ \emph {et~al.}(2022)\citenamefont {Haque}, \citenamefont {McClarty},\ and\ \citenamefont {Khaymovich}}]{Haque2022}%
  \BibitemOpen
  \bibfield  {author} {\bibinfo {author} {\bibfnamefont {M.}~\bibnamefont {Haque}}, \bibinfo {author} {\bibfnamefont {P.~A.}\ \bibnamefont {McClarty}},\ and\ \bibinfo {author} {\bibfnamefont {I.~M.}\ \bibnamefont {Khaymovich}},\ }\bibfield  {title} {\bibinfo {title} {Entanglement of midspectrum eigenstates of chaotic many-body systems: Reasons for deviation from random ensembles},\ }\href {https://doi.org/10.1103/PhysRevE.105.014109} {\bibfield  {journal} {\bibinfo  {journal} {Phys. Rev. E}\ }\textbf {\bibinfo {volume} {105}},\ \bibinfo {pages} {014109} (\bibinfo {year} {2022})}\BibitemShut {NoStop}%
\bibitem [{\citenamefont {Roberts}\ \emph {et~al.}(2018)\citenamefont {Roberts}, \citenamefont {Stanford},\ and\ \citenamefont {Streicher}}]{Roberts2018}%
  \BibitemOpen
  \bibfield  {author} {\bibinfo {author} {\bibfnamefont {D.~A.}\ \bibnamefont {Roberts}}, \bibinfo {author} {\bibfnamefont {D.}~\bibnamefont {Stanford}},\ and\ \bibinfo {author} {\bibfnamefont {A.}~\bibnamefont {Streicher}},\ }\bibfield  {title} {\bibinfo {title} {Operator growth in the syk model},\ }\href {https://doi.org/10.1007/JHEP06(2018)122} {\bibfield  {journal} {\bibinfo  {journal} {Journal of High Energy Physics}\ }\textbf {\bibinfo {volume} {2018}},\ \bibinfo {pages} {1} (\bibinfo {year} {2018})}\BibitemShut {NoStop}%
\bibitem [{\citenamefont {Parker}\ \emph {et~al.}(2019)\citenamefont {Parker}, \citenamefont {Cao}, \citenamefont {Avdoshkin}, \citenamefont {Scaffidi},\ and\ \citenamefont {Altman}}]{Parker2019}%
  \BibitemOpen
  \bibfield  {author} {\bibinfo {author} {\bibfnamefont {D.~E.}\ \bibnamefont {Parker}}, \bibinfo {author} {\bibfnamefont {X.}~\bibnamefont {Cao}}, \bibinfo {author} {\bibfnamefont {A.}~\bibnamefont {Avdoshkin}}, \bibinfo {author} {\bibfnamefont {T.}~\bibnamefont {Scaffidi}},\ and\ \bibinfo {author} {\bibfnamefont {E.}~\bibnamefont {Altman}},\ }\bibfield  {title} {\bibinfo {title} {A universal operator growth hypothesis},\ }\href {https://doi.org/10.1103/PhysRevX.9.041017} {\bibfield  {journal} {\bibinfo  {journal} {Phys. Rev. X}\ }\textbf {\bibinfo {volume} {9}},\ \bibinfo {pages} {041017} (\bibinfo {year} {2019})}\BibitemShut {NoStop}%
\bibitem [{\citenamefont {Zhuang}\ \emph {et~al.}(2019)\citenamefont {Zhuang}, \citenamefont {Schuster}, \citenamefont {Yoshida},\ and\ \citenamefont {Yao}}]{zhuang2019}%
  \BibitemOpen
  \bibfield  {author} {\bibinfo {author} {\bibfnamefont {Q.}~\bibnamefont {Zhuang}}, \bibinfo {author} {\bibfnamefont {T.}~\bibnamefont {Schuster}}, \bibinfo {author} {\bibfnamefont {B.}~\bibnamefont {Yoshida}},\ and\ \bibinfo {author} {\bibfnamefont {N.~Y.}\ \bibnamefont {Yao}},\ }\bibfield  {title} {\bibinfo {title} {Scrambling and complexity in phase space},\ }\href {https://doi.org/10.1103/PhysRevA.99.062334} {\bibfield  {journal} {\bibinfo  {journal} {Phys. Rev. A}\ }\textbf {\bibinfo {volume} {99}},\ \bibinfo {pages} {062334} (\bibinfo {year} {2019})}\BibitemShut {NoStop}%
\bibitem [{\citenamefont {Omanakuttan}\ \emph {et~al.}(2023)\citenamefont {Omanakuttan}, \citenamefont {Chinni}, \citenamefont {Blocher},\ and\ \citenamefont {Poggi}}]{omanakuttan2023}%
  \BibitemOpen
  \bibfield  {author} {\bibinfo {author} {\bibfnamefont {S.}~\bibnamefont {Omanakuttan}}, \bibinfo {author} {\bibfnamefont {K.}~\bibnamefont {Chinni}}, \bibinfo {author} {\bibfnamefont {P.~D.}\ \bibnamefont {Blocher}},\ and\ \bibinfo {author} {\bibfnamefont {P.~M.}\ \bibnamefont {Poggi}},\ }\bibfield  {title} {\bibinfo {title} {Scrambling and quantum chaos indicators from long-time properties of operator distributions},\ }\href@noop {} {\bibfield  {journal} {\bibinfo  {journal} {Physical Review A}\ }\textbf {\bibinfo {volume} {107}},\ \bibinfo {pages} {032418} (\bibinfo {year} {2023})}\BibitemShut {NoStop}%
\bibitem [{\citenamefont {Schuster}\ and\ \citenamefont {Yao}(2023)}]{schuster2023}%
  \BibitemOpen
  \bibfield  {author} {\bibinfo {author} {\bibfnamefont {T.}~\bibnamefont {Schuster}}\ and\ \bibinfo {author} {\bibfnamefont {N.~Y.}\ \bibnamefont {Yao}},\ }\bibfield  {title} {\bibinfo {title} {Operator growth in open quantum systems},\ }\href@noop {} {\bibfield  {journal} {\bibinfo  {journal} {Physical Review Letters}\ }\textbf {\bibinfo {volume} {131}},\ \bibinfo {pages} {160402} (\bibinfo {year} {2023})}\BibitemShut {NoStop}%
\bibitem [{\citenamefont {Swingle}\ \emph {et~al.}(2016)\citenamefont {Swingle}, \citenamefont {Bentsen}, \citenamefont {Schleier-Smith},\ and\ \citenamefont {Hayden}}]{swingle2016}%
  \BibitemOpen
  \bibfield  {author} {\bibinfo {author} {\bibfnamefont {B.}~\bibnamefont {Swingle}}, \bibinfo {author} {\bibfnamefont {G.}~\bibnamefont {Bentsen}}, \bibinfo {author} {\bibfnamefont {M.}~\bibnamefont {Schleier-Smith}},\ and\ \bibinfo {author} {\bibfnamefont {P.}~\bibnamefont {Hayden}},\ }\bibfield  {title} {\bibinfo {title} {Measuring the scrambling of quantum information},\ }\href@noop {} {\bibfield  {journal} {\bibinfo  {journal} {Physical Review A}\ }\textbf {\bibinfo {volume} {94}},\ \bibinfo {pages} {040302} (\bibinfo {year} {2016})}\BibitemShut {NoStop}%
\bibitem [{\citenamefont {G{\"a}rttner}\ \emph {et~al.}(2017)\citenamefont {G{\"a}rttner}, \citenamefont {Bohnet}, \citenamefont {Safavi-Naini}, \citenamefont {Wall}, \citenamefont {Bollinger},\ and\ \citenamefont {Rey}}]{garttner2017}%
  \BibitemOpen
  \bibfield  {author} {\bibinfo {author} {\bibfnamefont {M.}~\bibnamefont {G{\"a}rttner}}, \bibinfo {author} {\bibfnamefont {J.~G.}\ \bibnamefont {Bohnet}}, \bibinfo {author} {\bibfnamefont {A.}~\bibnamefont {Safavi-Naini}}, \bibinfo {author} {\bibfnamefont {M.~L.}\ \bibnamefont {Wall}}, \bibinfo {author} {\bibfnamefont {J.~J.}\ \bibnamefont {Bollinger}},\ and\ \bibinfo {author} {\bibfnamefont {A.~M.}\ \bibnamefont {Rey}},\ }\bibfield  {title} {\bibinfo {title} {Measuring out-of-time-order correlations and multiple quantum spectra in a trapped-ion quantum magnet},\ }\href@noop {} {\bibfield  {journal} {\bibinfo  {journal} {Nature Physics}\ }\textbf {\bibinfo {volume} {13}},\ \bibinfo {pages} {781} (\bibinfo {year} {2017})}\BibitemShut {NoStop}%
\bibitem [{\citenamefont {Vermersch}\ \emph {et~al.}(2019)\citenamefont {Vermersch}, \citenamefont {Elben}, \citenamefont {Sieberer}, \citenamefont {Yao},\ and\ \citenamefont {Zoller}}]{vermersch2019}%
  \BibitemOpen
  \bibfield  {author} {\bibinfo {author} {\bibfnamefont {B.}~\bibnamefont {Vermersch}}, \bibinfo {author} {\bibfnamefont {A.}~\bibnamefont {Elben}}, \bibinfo {author} {\bibfnamefont {L.~M.}\ \bibnamefont {Sieberer}}, \bibinfo {author} {\bibfnamefont {N.~Y.}\ \bibnamefont {Yao}},\ and\ \bibinfo {author} {\bibfnamefont {P.}~\bibnamefont {Zoller}},\ }\bibfield  {title} {\bibinfo {title} {Probing scrambling using statistical correlations between randomized measurements},\ }\href@noop {} {\bibfield  {journal} {\bibinfo  {journal} {Physical Review X}\ }\textbf {\bibinfo {volume} {9}},\ \bibinfo {pages} {021061} (\bibinfo {year} {2019})}\BibitemShut {NoStop}%
\bibitem [{\citenamefont {Landsman}\ \emph {et~al.}(2019)\citenamefont {Landsman}, \citenamefont {Figgatt}, \citenamefont {Schuster}, \citenamefont {Linke}, \citenamefont {Yoshida}, \citenamefont {Yao},\ and\ \citenamefont {Monroe}}]{landsman2019}%
  \BibitemOpen
  \bibfield  {author} {\bibinfo {author} {\bibfnamefont {K.~A.}\ \bibnamefont {Landsman}}, \bibinfo {author} {\bibfnamefont {C.}~\bibnamefont {Figgatt}}, \bibinfo {author} {\bibfnamefont {T.}~\bibnamefont {Schuster}}, \bibinfo {author} {\bibfnamefont {N.~M.}\ \bibnamefont {Linke}}, \bibinfo {author} {\bibfnamefont {B.}~\bibnamefont {Yoshida}}, \bibinfo {author} {\bibfnamefont {N.~Y.}\ \bibnamefont {Yao}},\ and\ \bibinfo {author} {\bibfnamefont {C.}~\bibnamefont {Monroe}},\ }\bibfield  {title} {\bibinfo {title} {Verified quantum information scrambling},\ }\href@noop {} {\bibfield  {journal} {\bibinfo  {journal} {Nature}\ }\textbf {\bibinfo {volume} {567}},\ \bibinfo {pages} {61} (\bibinfo {year} {2019})}\BibitemShut {NoStop}%
\bibitem [{\citenamefont {Mi}\ \emph {et~al.}(2021)\citenamefont {Mi}, \citenamefont {Roushan}, \citenamefont {Quintana}, \citenamefont {Mandra}, \citenamefont {Marshall}, \citenamefont {Neill}, \citenamefont {Arute}, \citenamefont {Arya}, \citenamefont {Atalaya}, \citenamefont {Babbush} \emph {et~al.}}]{scr_scqubits2021}%
  \BibitemOpen
  \bibfield  {author} {\bibinfo {author} {\bibfnamefont {X.}~\bibnamefont {Mi}}, \bibinfo {author} {\bibfnamefont {P.}~\bibnamefont {Roushan}}, \bibinfo {author} {\bibfnamefont {C.}~\bibnamefont {Quintana}}, \bibinfo {author} {\bibfnamefont {S.}~\bibnamefont {Mandra}}, \bibinfo {author} {\bibfnamefont {J.}~\bibnamefont {Marshall}}, \bibinfo {author} {\bibfnamefont {C.}~\bibnamefont {Neill}}, \bibinfo {author} {\bibfnamefont {F.}~\bibnamefont {Arute}}, \bibinfo {author} {\bibfnamefont {K.}~\bibnamefont {Arya}}, \bibinfo {author} {\bibfnamefont {J.}~\bibnamefont {Atalaya}}, \bibinfo {author} {\bibfnamefont {R.}~\bibnamefont {Babbush}}, \emph {et~al.},\ }\bibfield  {title} {\bibinfo {title} {Information scrambling in quantum circuits},\ }\href@noop {} {\bibfield  {journal} {\bibinfo  {journal} {Science}\ }\textbf {\bibinfo {volume} {374}},\ \bibinfo {pages} {1479} (\bibinfo {year} {2021})}\BibitemShut {NoStop}%
\bibitem [{\citenamefont {Joshi}\ \emph {et~al.}(2020)\citenamefont {Joshi}, \citenamefont {Elben}, \citenamefont {Vermersch}, \citenamefont {Brydges}, \citenamefont {Maier}, \citenamefont {Zoller}, \citenamefont {Blatt},\ and\ \citenamefont {Roos}}]{scr_ions2020}%
  \BibitemOpen
  \bibfield  {author} {\bibinfo {author} {\bibfnamefont {M.~K.}\ \bibnamefont {Joshi}}, \bibinfo {author} {\bibfnamefont {A.}~\bibnamefont {Elben}}, \bibinfo {author} {\bibfnamefont {B.}~\bibnamefont {Vermersch}}, \bibinfo {author} {\bibfnamefont {T.}~\bibnamefont {Brydges}}, \bibinfo {author} {\bibfnamefont {C.}~\bibnamefont {Maier}}, \bibinfo {author} {\bibfnamefont {P.}~\bibnamefont {Zoller}}, \bibinfo {author} {\bibfnamefont {R.}~\bibnamefont {Blatt}},\ and\ \bibinfo {author} {\bibfnamefont {C.~F.}\ \bibnamefont {Roos}},\ }\bibfield  {title} {\bibinfo {title} {Quantum information scrambling in a trapped-ion quantum simulator with tunable range interactions},\ }\href@noop {} {\bibfield  {journal} {\bibinfo  {journal} {Physical Review Letters}\ }\textbf {\bibinfo {volume} {124}},\ \bibinfo {pages} {240505} (\bibinfo {year} {2020})}\BibitemShut {NoStop}%
\bibitem [{\citenamefont {Liang}\ \emph {et~al.}(2024)\citenamefont {Liang}, \citenamefont {Yue}, \citenamefont {Chao}, \citenamefont {Hua}, \citenamefont {Lin}, \citenamefont {Tey},\ and\ \citenamefont {You}}]{scr_rydberg2024}%
  \BibitemOpen
  \bibfield  {author} {\bibinfo {author} {\bibfnamefont {X.}~\bibnamefont {Liang}}, \bibinfo {author} {\bibfnamefont {Z.}~\bibnamefont {Yue}}, \bibinfo {author} {\bibfnamefont {Y.-X.}\ \bibnamefont {Chao}}, \bibinfo {author} {\bibfnamefont {Z.-X.}\ \bibnamefont {Hua}}, \bibinfo {author} {\bibfnamefont {Y.}~\bibnamefont {Lin}}, \bibinfo {author} {\bibfnamefont {M.~K.}\ \bibnamefont {Tey}},\ and\ \bibinfo {author} {\bibfnamefont {L.}~\bibnamefont {You}},\ }\bibfield  {title} {\bibinfo {title} {Observation of anomalous information scrambling in a rydberg atom array},\ }\href@noop {} {\bibfield  {journal} {\bibinfo  {journal} {arXiv preprint arXiv:2410.16174}\ } (\bibinfo {year} {2024})}\BibitemShut {NoStop}%
\bibitem [{\citenamefont {Srednicki}(1994)}]{Srednicki1994}%
  \BibitemOpen
  \bibfield  {author} {\bibinfo {author} {\bibfnamefont {M.}~\bibnamefont {Srednicki}},\ }\bibfield  {title} {\bibinfo {title} {Chaos and quantum thermalization},\ }\href {https://doi.org/10.1103/PhysRevE.50.888} {\bibfield  {journal} {\bibinfo  {journal} {Phys. Rev. E}\ }\textbf {\bibinfo {volume} {50}},\ \bibinfo {pages} {888} (\bibinfo {year} {1994})}\BibitemShut {NoStop}%
\bibitem [{\citenamefont {Deutsch}(1991)}]{Deutsch1991}%
  \BibitemOpen
  \bibfield  {author} {\bibinfo {author} {\bibfnamefont {J.~M.}\ \bibnamefont {Deutsch}},\ }\bibfield  {title} {\bibinfo {title} {Quantum statistical mechanics in a closed system},\ }\href {https://doi.org/10.1103/PhysRevA.43.2046} {\bibfield  {journal} {\bibinfo  {journal} {Phys. Rev. A}\ }\textbf {\bibinfo {volume} {43}},\ \bibinfo {pages} {2046} (\bibinfo {year} {1991})}\BibitemShut {NoStop}%
\bibitem [{\citenamefont {Lerose}\ \emph {et~al.}(2023)\citenamefont {Lerose}, \citenamefont {Parolini}, \citenamefont {Fazio}, \citenamefont {Abanin},\ and\ \citenamefont {Pappalardi}}]{lerose2023}%
  \BibitemOpen
  \bibfield  {author} {\bibinfo {author} {\bibfnamefont {A.}~\bibnamefont {Lerose}}, \bibinfo {author} {\bibfnamefont {T.}~\bibnamefont {Parolini}}, \bibinfo {author} {\bibfnamefont {R.}~\bibnamefont {Fazio}}, \bibinfo {author} {\bibfnamefont {D.~A.}\ \bibnamefont {Abanin}},\ and\ \bibinfo {author} {\bibfnamefont {S.}~\bibnamefont {Pappalardi}},\ }\bibfield  {title} {\bibinfo {title} {Theory of robust quantum many-body scars in long-range interacting systems},\ }\href@noop {} {\bibfield  {journal} {\bibinfo  {journal} {arXiv preprint arXiv:2309.12504}\ } (\bibinfo {year} {2023})}\BibitemShut {NoStop}%
\bibitem [{\citenamefont {Pandey}\ \emph {et~al.}(2020)\citenamefont {Pandey}, \citenamefont {Claeys}, \citenamefont {Campbell}, \citenamefont {Polkovnikov},\ and\ \citenamefont {Sels}}]{pandey2020}%
  \BibitemOpen
  \bibfield  {author} {\bibinfo {author} {\bibfnamefont {M.}~\bibnamefont {Pandey}}, \bibinfo {author} {\bibfnamefont {P.~W.}\ \bibnamefont {Claeys}}, \bibinfo {author} {\bibfnamefont {D.~K.}\ \bibnamefont {Campbell}}, \bibinfo {author} {\bibfnamefont {A.}~\bibnamefont {Polkovnikov}},\ and\ \bibinfo {author} {\bibfnamefont {D.}~\bibnamefont {Sels}},\ }\bibfield  {title} {\bibinfo {title} {Adiabatic eigenstate deformations as a sensitive probe for quantum chaos},\ }\href@noop {} {\bibfield  {journal} {\bibinfo  {journal} {Physical Review X}\ }\textbf {\bibinfo {volume} {10}},\ \bibinfo {pages} {041017} (\bibinfo {year} {2020})}\BibitemShut {NoStop}%
\bibitem [{\citenamefont {Varma}\ \emph {et~al.}(2024)\citenamefont {Varma}, \citenamefont {Vardi},\ and\ \citenamefont {Cohen}}]{varma2024}%
  \BibitemOpen
  \bibfield  {author} {\bibinfo {author} {\bibfnamefont {A.~V.}\ \bibnamefont {Varma}}, \bibinfo {author} {\bibfnamefont {A.}~\bibnamefont {Vardi}},\ and\ \bibinfo {author} {\bibfnamefont {D.}~\bibnamefont {Cohen}},\ }\bibfield  {title} {\bibinfo {title} {Many-body adiabatic passage: Instability, chaos, and quantum classical correspondence},\ }\href@noop {} {\bibfield  {journal} {\bibinfo  {journal} {arXiv preprint arXiv:2409.00952}\ } (\bibinfo {year} {2024})}\BibitemShut {NoStop}%
\bibitem [{\citenamefont {Kumari}\ and\ \citenamefont {Alhambra}(2022)}]{Kumari2022}%
  \BibitemOpen
  \bibfield  {author} {\bibinfo {author} {\bibfnamefont {M.}~\bibnamefont {Kumari}}\ and\ \bibinfo {author} {\bibfnamefont {{\'{A}}.~M.}\ \bibnamefont {Alhambra}},\ }\bibfield  {title} {\bibinfo {title} {Eigenstate entanglement in integrable collective spin models},\ }\href {https://doi.org/10.22331/q-2022-04-27-701} {\bibfield  {journal} {\bibinfo  {journal} {{Quantum}}\ }\textbf {\bibinfo {volume} {6}},\ \bibinfo {pages} {701} (\bibinfo {year} {2022})}\BibitemShut {NoStop}%
\end{thebibliography}%


%

\end{document}